\documentclass[12pt]{article}
\usepackage{amssymb}
\usepackage{eqsection,latexsym,epsf}
\usepackage{epsfig,amsfonts,cite}

\def\zh{\hat{\zeta}}

\def\tb{t_{\beta}}
\def\tx{{\tt x}}
\def\utx{\underline{\tx}}
\def\t0{\tt 0}
\def\ut0{\underline{\tt 0}}
\def\l{\lambda}
\def\lh{\widehat{\lambda}}
\def\vs{\vec{\sigma}}

\def\us{\underline{\sigma}}
\def\fh{\widehat{f}}
\def\rh{\hat{\rho}}
\def\Qh{\widehat{Q}}
\def\ph{\widehat{\pi}}
\def\la{\overline{l}}
\def\ka{\overline{k}}
\def\atanh{{\rm arctanh}}
\def\uh{\hat{u}}

\newcommand{\be}{\begin{equation}}
\newcommand{\ee}{\end{equation}}
\newcommand{\bea}{\begin{eqnarray}}
\newcommand{\eea}{\end{eqnarray}}
\newcommand{\<}{\langle}
\renewcommand{\>}{\rangle}

\begin{document}

\title{The Dynamic Phase Transition for Decoding Algorithms}

\author{       
  { Silvio Franz}              \\
  {\small\it International Center for Theoretical Physics}
  \\[-0.2cm]
  {\small\it P.O. Box 586, I-34100 Trieste, ITALY}        \\[-0.2cm]
  {\small Internet: {\tt franz,micleone@ictp.trieste.it}}
          \\[0.5cm]
  { Michele Leone}              \\
  {\small\it International Center for Theoretical Physics and SISSA}
  \\[-0.2cm]
  {\small\it via Beirut 8, I-34100 Trieste, ITALY}        \\[-0.2cm]
  {\small Internet: {\tt franz,micleone@ictp.trieste.it}}
          \\[0.5cm]
  { Andrea Montanari}              \\
  {\small\it Laboratoire de Physique Th\'{e}orique de l'Ecole Normale
  Sup\'{e}rieure\footnote {UMR 8549, Unit{\'e}   Mixte de Recherche du 
Centre National de la Recherche Scientifique et de 
l' Ecole Normale Sup{\'e}rieure. } }
  \\[-0.2cm]
  {\small\it 24, rue Lhomond, 75231 Paris CEDEX 05, FRANCE}        \\[-0.2cm]
  {\small Internet: {\tt Andrea.Montanari@lpt.ens.fr}}
          \\[0.5cm]
  { Federico Ricci-Tersenghi}              \\
  {\small\it Dipartimento di Fisica and SMC and UdR1 of INFM}\\[-0.2cm]
  {\small\it Universit\`a di Roma "La Sapienza"}\\[-0.2cm]
  {\small\it Piazzale Aldo Moro 2, I-00185 Roma, ITALY}\\[-0.2cm]
  {\small Internet: {\tt Federico.Ricci@roma1.infn.it}}
          \\[-0.1cm]
  {\protect\makebox[5in]{\quad}}  % To force authors' names to be written
                                  %   vertically, one above another.
                                  % (\author seems to put them side-by-side
                                  %   if there is room.)
%  \\
}

\maketitle
\vspace{-17.cm}
\begin{flushright} 
LPTENS 02/31   
\end{flushright}
\vspace{17.cm}

\thispagestyle{empty} 

\abstract{The state-of-the-art error correcting codes are based on large
random constructions (random graphs, random permutations, \dots)
and are decoded by linear-time iterative algorithms. 
Because of these features, they are remarkable examples of 
diluted mean-field spin glasses, both
from the static and from the dynamic points of view.
We analyze the behavior of decoding algorithms using the
mapping onto statistical-physics models. This allows
to understand the intrinsic (i.e. algorithm independent) features 
of this behavior.}

\clearpage

%
%*********************************************************************
%
\section{Introduction}

Recently there has been some interest in studying 
``complexity phase transitions'', i.e. abrupt changes in the 
computational complexity of hard combinatorial problems as some control 
parameter is varied \cite{CSReview}. 
These phenomena are thought to be somehow related 
to the physics of glassy systems, where the physical dynamics
experiences a dramatic slowing down as the temperature is 
lowered \cite{DynamicsReview}.

Complexity is a central issue also in coding theory 
\cite{BargComplexity,SpielmanLecture}. 
Coding theory \cite{Cover,GallagerBook,Viterbi}
deals with the problem of communicating information 
reliably through an unreliable channel of communication. 
This task is accomplished by making use of 
{\it error correcting codes}. In 1948 Shannon \cite{Shannon}
proved that almost any 
error correcting code allows to communicate without errors, as long
as the rate of transmitted information is kept below the {\it capacity} of the
channel. However decoding is an intractable problem for almost any code.
Coding theory is therefore a rich source of interesting computational
problems. 

On the other hand it is known that error correcting codes can be mapped
onto disordered spin models \cite{Sourlas1,Sourlas2,Sourlas4,Rujan,Sourlas5}. 
Remarkably there has recently been 
a revolution in coding theory which has brought to the invention 
of new and very powerful codes based on random constructions:
turbo codes \cite{PrimoBerrou}, 
low density parity check codes (LDPCC) \cite{GallagerThesis,MacKay}, 
repetition accumulated codes \cite{RA}, etc. 
As a matter of fact the equivalent spin models
have been intensively studied in the last few years.
These are diluted spin glasses, i.e. spin glasses on random 
(hyper)graphs \cite{MonassonRSB,MezardParisiBethe,FranzEtAlExact,FranzEtAlFerromagn}.

The new codes are decoded by using approximate iterative
algorithms, which are closely related to the TAP-cavity approach to
mean field spin glasses \cite{YedidiaFirst,YedidiaLast}.
We think therefore that a close investigation of these systems 
from a statistical physics point of view, having in mind complexity
(i.e. dynamical) issues, can be of great theoretical 
interest\footnote{The reader is urged to consult Refs. 
\cite{KanterSaad_PRL,SaadRegular,KanterSaad_Cascading,Turbo1,Turbo2,
Saad_MN_PRL,KanterSaad_FS,SaadCactus,GallagerAM,SaadTighter} 
for a statistical mechanics analysis of the optimal decoding (i.e.
of static issues).}.

\begin{figure}
\centerline{\hspace{0.cm}
\epsfig{figure=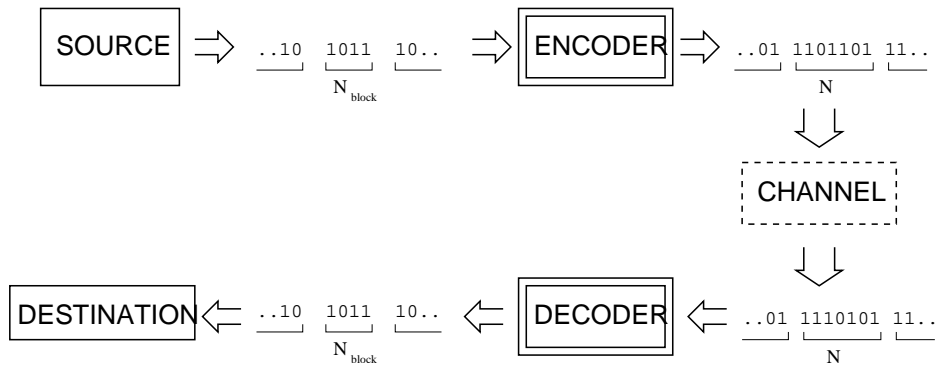,angle=0,width=0.7\linewidth}}
\caption{A schematic description of how error correcting codes work.}
\label{ChannelFig}
\end{figure}
Let us briefly recall the general setting of coding theory \cite{Cover}
in order to fix a few notations (cf. Fig. \ref{ChannelFig}
for a pictorial description). 
A source of information produces a stream of symbols.
Let us assume, for instance, that the source produces unbiased random bits. 
The stream is partitioned into {\it blocks} of
length $N_{\rm block}$. Each of the possible $2^{N_{\rm block}}$ blocks 
is mapped to a {\it codeword} (i.e. a sequence of bits)
of length $N>N_{\rm block}$ by the {\it encoder} and transmitted 
through the channel. An error correcting code is therefore
defined either as a mapping 
$\{{\tt 0},{\tt 1}\}^{N_{\rm block}}\to\{{\tt 0},{\tt 1}\}^{N}$, or
as a list of $2^{N_{\rm block}}$ codewords.
The {\it rate} of the code is defined as $R=N_{\rm block}/N$.

Let us  denote\footnote{We shall denote transmitted and received
symbols by typographic characters, with the exception of symbols in 
$\{+1,-1\}$. In this case use the physicists notation and denote 
such symbols by $\sigma$. When considering binary symbols we will
often pass from the $\tx$ notation to the $\sigma$ notation, the 
correspondence $\sigma=(-1)^{\tx}$ being understood.
Finally vectors of length $N$ will be always
denoted by underlined characters: e.g. $\utx$ or $\us$.} 
the transmitted codeword by
$\utx^{\rm in} =[\tx^{\rm in}_1,\dots,\tx^{\rm in}_N]^{\tt T}$.
Due to the noise, a different sequence of symbols 
$\utx^{\rm out} =[\tx^{\rm out}_1,\dots,\tx^{\rm out}_N]^{\tt T}$
is received. The decoding problem is to infer $\utx^{\rm in}$ given
$\utx^{\rm out}$, the definition of the code, and the properties of the 
noisy channel.

It is useful to summarize the general picture which emerges from our work.
We shall focus on Gallager codes (both {\it regular} and {\it irregular}).
The optimal decoding strategy (maximum-likelihood decoding) is able
to recover the transmitted message below some noise threshold: $p<p_c$.
Iterative, linear time, algorithms get stuck (in general)
at a lower noise level, and are successful only for $p<p_d({\rm alg.})$,
with $p_d({\rm alg.})\le p_c$. 
In general the ``dynamical'' threshold $p_d({\rm alg.})$ 
depends upon the details of the algorithm.  However, it seems to be always
smaller than  some universal (although code-dependent) value $p_d$.
Moreover, some ``optimal'' linear-time algorithms are successful up
to $p_d$ (i.e. $p_d({\rm alg.})=p_d$). The universal threshold $p_d$
coincides with the dynamical transition \cite{DynamicsReview} 
of the corresponding spin model.

The plan of the paper is the following. In Section 
\ref{CodingDecodingSection} we introduce low density parity check codes 
(LDPCC), focusing on Gallager's {\it ensembles}, and we describe
{\it message passing} decoding algorithms. We briefly recall the connection
between this algorithms and the TAP-cavity equations for mean-field
spin glasses. In Sec. \ref{ReplicaSection} we define
a spin model which describes the decoding problem, and introduce the replica 
formalism. In Sec. \ref{BECSection} we analyze this model for a 
particular choice of the noisy channel (the {\it binary erasure channel}).
In this case calculations can be fully explicit and the results are
particularly clear. Then, in Sec. \ref{GeneralChannelSection}, we 
address the general case. Finally we draw our conclusions in Sec. 
\ref{ConclusionSection}. The Appendices collect some details of 
our computations.
%
%*********************************************************************
%
\section{Error correcting codes, decoding algorithms and the cavity equations}
\label{CodingDecodingSection}

This Section introduces the reader to some basic terminology in coding theory.
In the first part we define some {\it ensembles} of codes, namely 
{\it regular} and {\it irregular} LDPCC. In the second one we describe
a class of iterative decoding algorithms. These algorithms have 
a very clear physical interpretation, which we briefly recall.
Finally we explain how these algorithms are analyzed in the coding theory
community.
This Section does not contain any original result. The interested reader
may consult Refs. 
\cite{RichardsonUrbankeIntroduction,GallagerThesis,GallagerBook,YedidiaLast} 
for further details.

\subsection{Encoding $\dots$}
\label{EncodingExplanation}

Low density parity check codes are defined by assigning a 
binary $N\times M$ matrix ${\mathbb H} =\{H_{ij}\}$, with $H_{ij}\in\{0,1\}$.
All the codewords are required to satisfy the constraint
\begin{eqnarray}
{\mathbb H}\, \utx = 0\;\;\;\; ({\rm mod}\;\;\; 2)\, .
\label{ParityCheckMatrix}
\end{eqnarray}
The matrix ${\mathbb H}$ is called the {\it parity check matrix} and the 
$M$ equations summarized in Eq. (\ref{ParityCheckMatrix}) are the 
{\it parity check equations} (or, for short, {\it parity checks}).
If the matrix ${\mathbb H}$ has rank $M$ (this is usually the case), the
rate is $R=1-M/N$.

There exists a nice graphic representation of Eq. (\ref{ParityCheckMatrix})
which is often used in the coding theory community: the {\it Tanner
graph} representation \cite{Tanner,ForneyGraph}. 
One constructs a bipartite graph by 
associating a left-hand node to each one of the $N$ variables,
and a right-hand node to each one of the $M$ parity checks. An edge 
is drawn between the {\it variable node} $i$
and the {\it parity check node} $\alpha$ if and only if the variable $\tx_i$
appears with a non-zero coefficient in the parity check equation $\alpha$.

\begin{figure}
\centerline{\hspace{0.cm}
\epsfig{figure=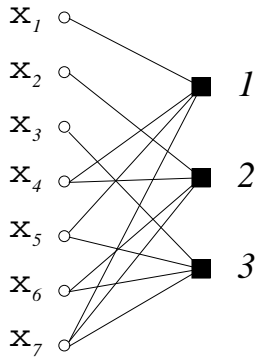,angle=0,width=0.2\linewidth}}
\caption{The Tanner graph for the ${\cal H}_2(3)$ Hamming code.}
\label{HammingFig}
\end{figure}
Let us for instance consider the celebrated ${\cal H}_2(3)$ Hamming code (one
of the first examples in any book on coding theory). In this case we have 
$N=7$, $M=3$ and 
\begin{eqnarray}
{\mathbb H} = \left[\begin{array}{ccccccc}
1 & 0 & 0 & 1 & 1 & 0 & 1 \\
0 & 1 & 0 & 1 & 0 & 1 & 1 \\
0 & 0 & 1 & 0 & 1 & 1 & 1 \\
\end{array}\right]\, .\label{HammingMatrix}
\end{eqnarray}
This code has $2^4=16$ codewords 
$\utx^{(\alpha)} = [\tx^{(\alpha)}_1,\dots,\tx^{(\alpha)}_7]^{\tt T}$,
with $\alpha \in\{1,\dots,16\}$.
They are the solutions of the three parity check equations:
$\tx_1+\tx_4+\tx_5+\tx_7 =0$; $\tx_2+\tx_4+\tx_6+\tx_7 =0$; 
$\tx_3+\tx_5+\tx_6+\tx_7 =0$ $({\rm mod} 2)$.
The corresponding Tanner graph is drawn in Fig. \ref{HammingFig}.

In general one considers {\it ensembles} of codes, by defining a random
construction of the parity check matrix. 
One of the simplest {\it ensembles} is given by {\it regular} $(k,l)$
{\it Gallager codes}. In this case one chooses the matrix ${\mathbb H}$
randomly among all the $N\times M$ matrices having $k$ non-zero
entries per row, and $l$ per column.
The Tanner graph is therefore a random bipartite graph with 
fixed degrees $k$ and $l$ respectively for the parity check nodes
and for the variable nodes. Of course this is possible only
if $M/N = l/k$.

Amazingly good codes \cite{Tornado,ImprovedLDPCC,Chung} 
where obtained by slightly more sophisticated 
{\it irregular} constructions.
In this case one assigns the distributions of the degrees of parity
check nodes and variable nodes in the Tanner graph.
We shall denote by $\{ c_k\}$ the degree distribution
of the check nodes and $\{ v_l\}$  the degree distribution of the
variable nodes. This means that there are $N v_l$ bits of
the codeword belonging to $l$ parity checks and $N c_k$ parity
checks involving $k$ bits for each $k$ and $l$.
We shall always assume $c_k=0$ for $k<3$ and $v_l=0$ for $l<2$

It is useful to define the generating polynomials
\begin{eqnarray}
c(x) \equiv\sum_{k=3}^{\infty}c_k x^k\, ,\;\;\;
v(x) \equiv \sum_{l=2}^{\infty}v_l x^l\, ,
\end{eqnarray}
which satisfy the normalization condition $c(1) = v(1) =1$.
Moreover we define the average variable and check degrees
$\overline{l} = v'(1)$ and $\overline{k} = c'(1)$.
Particular examples of this formalism are the regular codes,
whose generating polynomials are  $c(x) = x^k$,
$v(x) = x^l$.

\subsection{$\dots$ and decoding}

\begin{figure}
\centerline{\hspace{0.cm}
\epsfig{figure=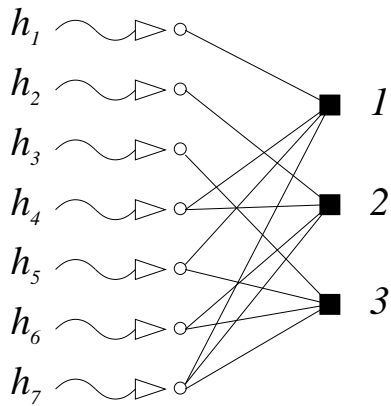,angle=0,width=0.3\linewidth}}
\caption{The information coming from the channel 
must be used for decoding the ${\cal H}_2(3)$ Hamming code: 
a pictorial view.}
\label{FieldFig}
\end{figure}

The codewords are transmitted trough a noisy channel.
We assume antipodal signalling: one sends $\sigma^{\rm in}\in\{+ 1, -1\}$ 
signals instead of $\tx^{\rm in}\in \{0,1\}$ through the channel
(the correspondence being given by $\sigma = (-1)^\tx$). 
At the end of the channel, a corrupted version of this signals is received. 
This means that  if $\sigma^{\rm in}\in\{+1,-1\}$ is transmitted, 
the value $\tx^{\rm out}$ is received
with probability density $Q(\tx^{\rm out}|\sigma^{\rm in})$.
The information conveyed by the received signal $\tx^{\rm out}$ is 
conveniently described by the log-likelihood\footnote{Notice the 
unconventional normalization: the factor $1/2$ is inserted to make 
contact with the statistical mechanics formulation.}:
\begin{eqnarray}
h(\tx^{\rm out}) = \frac{1}{2}
\log\frac{Q(\tx^{\rm out}|+1)}{Q(\tx^{\rm out}|-1)}\, .\label{Likelihood}
\end{eqnarray}
We can represent this information by wavy lines in the Tanner graph, 
cf. Fig. \ref{FieldFig}.

The decoding problem is to compute the probability for each transmitted bit 
$\sigma_i^{\rm in}$ to take the value $\sigma_i$, given the structure of the
code and the received message $\utx^{\rm out} = [\tx^{\rm out}_1,\dots ,
\tx^{\rm out}_N]^{\tt T}$.
This is in general an intractable problem 
\cite{BargComplexity,SpielmanLecture}. Recently there has been a 
great interest in dealing with this problem using approximate 
{\it message passing} algorithms. 

\begin{figure}
\centerline{\hspace{0cm}
\epsfig{figure=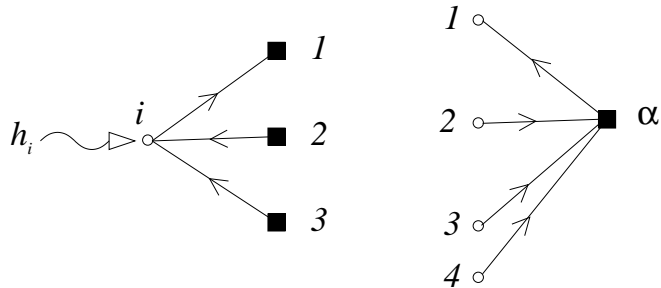,angle=0,width=0.5\linewidth}}
\caption{A graphic representation of the operations executed in a message
passing algorithm. At the variable node $i$ (on the left): 
$x^{(t+1)}_{i\to 1}=F(y^{(t)}_{2\to i},y^{(t)}_{3\to i};h_i)$.
At the check node $\alpha$ (on the right): 
$y^{(t+1)}_{\alpha\to 1}=G(x^{(t)}_{2\to \alpha},x^{(t)}_{3\to \alpha},
x^{(t)}_{4\to \alpha})$.}
\label{ComputationFig}
\end{figure}

Message passing algorithms are iterative:
at each step $t$ one keeps track of $M\ka$ messages from the
variable nodes to the check nodes $\{ y^{(t)}_{\alpha\to i}\}$
and viceversa $\{ x^{(t)}_{i\to \alpha}\}$.
Messages can be thought to travel along the edges and 
computations to be executed at the nodes. A node computes the 
message to be sent along each one of the edges, using the messages 
received from the other (!) edges at the previous iteration (the 
variable nodes make also use of  the log-likelihoods $h(\tx_i^{\rm out})$),
cf. Fig. \ref{ComputationFig}.
At some point the iteration is stopped (there exists no general 
stopping criterion), and a choice for the bit $\sigma_i$ is taken 
using all the incoming messages (plus the log-likelihood $h(\tx_i^{\rm out})$).

The functions which define the ``new'' messages in terms of the 
``old'' ones, can be chosen to optimize the decoder performances.
A particularly interesting family is the following:
\begin{eqnarray}
x^{(t+1)}_{i\to\alpha} &=& h_i +\sum_{\alpha'\ni i:\, \alpha'\neq\alpha}
y^{(t)}_{\alpha'\to i}\,\label{BeliefPropagation1}\\
y^{(t+1)}_{\alpha\to i} & =&
\frac{1}{\zeta}\, \atanh \left[\prod_{j\in\alpha:\, j\neq i}
\tanh\zeta x^{(t)}_{j\to\alpha}\right]\, ,\label{BeliefPropagation2}
\end{eqnarray}
where we used the notation $i\in \alpha$ whenever the bit $i$ 
belongs to the parity check $\alpha$.
The messages $\{x^{(\cdot)}_{i\to\alpha}\}$ and 
$\{ y^{(\cdot)}_{\alpha\to i}\}$ can be rescaled in such a way to 
eliminate the parameter $\zeta$ everywhere except in front of 
$h_i$. Therefore $\zeta$ allows to tune the importance given to the
information contained in the received message.

After the convergence of the above iteration one computes 
the {\it a posteriori} log-likelihoods as follows:
\begin{eqnarray}
H_i = h_i +\sum_{\alpha\ni i}y^{(\infty)}_{\alpha\to i}\, .
\end{eqnarray}
The meaning of the $\{H_i\}$ is analogous to the one of the $\{ h_i\}$
(but for the fact that the $H_i$ incorporate the information coming 
from the structure of the code): the best guess for the bit $i$
is $\sigma_i=+1$ or $\sigma_i = -1$ depending whether $H_i>0$ 
or $H_i<0$. 

The most popular choice for the free parameter $\zeta$ 
is $\zeta=1$: this algorithm has been 
invented separately by R.~G.~Gallager \cite{GallagerThesis}
in the coding theory context (and named
the {\it sum-product} algorithm) and by D.~Pearl \cite{Pearl} 
in the artificial  intelligence context (and named the 
{\it belief propagation} algorithm). Also $\zeta = \infty$ is sometimes 
used (the {\it max-product} algorithm).

The alerted reader will notice that the Eqs. 
(\ref{BeliefPropagation1})-(\ref{BeliefPropagation2}) are nothing but the 
cavity equations at inverse temperature $\zeta$ for a properly constructed
spin model. This remark is the object of Refs. \cite{SaadTAP,YedidiaFirst}.

In the analysis of the above algorithm it is convenient to assume
that $\sigma^{\rm in}_i=+1$ for $i=1,\dots,N$. 
This assumption can be made without loss of generality if the channel
is symmetric (i.e. if $Q({\tt x}|+1)=Q(-{\tt x}|-1)$).
With this assumption the $h_i$ are i.i.d. random variables with density
\begin{eqnarray}
p(h) \equiv Q(\tx(h)|+1)|\tx'(h)|\, ,
\end{eqnarray}
where $\tx(h)$ is the function which inverts Eq. (\ref{Likelihood}).
In the following we shall consider two particular examples of noisy 
channels, the generalization being straightforward:
\begin{itemize}
\item The binary erasure channel (BEC).  In this case a bit can either 
be received correctly or erased\footnote{This is what happens, for instance,
to packets in the Internet traffic.}. 
There are therefore three possible outputs:
$\{+ 1, -1, 0\}$. The transition probability is:
\begin{eqnarray}
Q(\tx^{\rm out}|+1) = \left\{\begin{array}{rl} 
(1-p) & \mbox{if }\tx^{\rm out} = +1\, ,\\
p     & \mbox{if }\tx^{\rm out} = 0\, ,\\
0     & \mbox{if }\tx^{\rm out} = -1\, ,
\end{array}\right.\;\;\;
Q(\tx^{\rm out}|-1) = \left\{\begin{array}{rl} 
0 & \mbox{if }\tx^{\rm out} = +1\, ,\\
p     & \mbox{if }\tx^{\rm out} = 0\, ,\\
(1-p)     & \mbox{if }\tx^{\rm out} = -1\, . 
\end{array}\right.\nonumber\\
\end{eqnarray}
We get therefore the following distribution for the log-likelihoods:
 $p(h) = (1-p)\, \delta_{\infty}(h)+ p\, \delta(h)$ (where $\delta_{\infty}$
is a Dirac delta function centered at $+\infty$).
Let us recall that the capacity of the BEC is given by
$C_{BEC} = 1-p$: this means that a rate-$R$ code cannot assure 
error correction if $p>1-R$.
\item The binary symmetric channel (BSC). The channel flips
each bit independently with probability $p$.  Namely
\begin{eqnarray}
Q(\tx^{\rm out}|+1) = \left\{\begin{array}{rl} 
(1-p) & \mbox{if }\tx^{\rm out} = +1\, ,\\
p     & \mbox{if }\tx^{\rm out} = -1\, ,
\end{array}\right.\;\;\;
Q(\tx^{\rm out}|-1) = \left\{\begin{array}{rl} 
p & \mbox{if }\tx^{\rm out} = +1\, ,\\
(1-p)     & \mbox{if }\tx^{\rm out} = -1\, . 
\end{array}\right.\nonumber\\
\end{eqnarray}
The corresponding log-likelihood distribution is $p(h) = (1-p)\,\delta(h-h_0)
+p\, \delta(h+h_0)$, with $h_0 = \atanh(1-2p)$.
The capacity of the BSC is\footnote{We denote
by ${\tt h}(p)$ the binary entropy function 
${\tt h}(p)=-p\log_2 p-(1-p)\log_2(1-p)$. It is useful to define its
inverse: we denote by $\delta_{GV}(R)$ (the so-called Gilbert-Varshamov
distance) the smallest solution of ${\tt h}(\delta) = 1-R$.} 
$C_{BSC} = 1-{\tt h}(p)$: a rate-$R$ code cannot correct errors 
if $p>\delta_{GV}(R)$.
\end{itemize}

It is quite easy \cite{RichardsonUrbanke,RichardsonUrbankeIntroduction}
to write a recursive equations for the probability 
distributions of the messages $\pi_t(x)$ and $\ph_t(y)$:
\begin{eqnarray}
\pi_{t+1}(x) & = & \frac{1}{\la}\sum_{l=2}^{\infty}v_ll
\int\!\prod_{i=1}^{l-1}dy_i\,\ph_t(y_i)\,\int\!dh\,p(h)\;
\delta\left(x-h-\sum_{i=1}^{l-1}y_i\right)\, ,\label{DensityEvolution_1}\\
\ph_{t+1}(y) & = & \frac{1}{\ka}\sum_{k=3}^{\infty}c_kk
\int\!\prod_{i=1}^{k-1}dx_i\,\pi_t(x_i)\;
\delta\left(y-\frac{1}{\zeta}\, \atanh \left[\prod_{i=1}^{k-1}
\tanh\zeta x_i\right]\right)\, .\label{DensityEvolution_2}
\end{eqnarray}
These equations (usually called the {\it density evolution} equations) 
are correct for times  $t\ll \log N$ due to the fact 
that the Tanner graph is locally tree-like.
They allow therefore to predict whether, for a given {\it ensemble} of codes 
and noise level (recall that the noise level is hidden in $p(h)$) 
the algorithm is able to recover 
the transmitted codeword (for large $N$). If this is the case, 
the distributions $\pi_t(x)$ and $\ph_t(y)$ will concentrate on 
$x=y=+\infty$ as $t\to\infty$. In the opposite case the above iteration will
converge to some distribution supported on finite values of $x$ and $y$.
\begin{table}
\centerline{
\begin{tabular}{|c|c|c|c|c|c|c|}
\hline
  & \multicolumn{2}{c|}{BEC} & \multicolumn{4}{c|}{BSC}\\
\hline
$(k,l)$ & $p_c$ & $p_d$  & $p_c$ & $p_d(\zeta=1)$ & $p_d(\zeta=2)$
& $p_d(\zeta=\infty)$ \\
\hline
$(6,3)$  & $0.4882$ & $0.4294$ & $0.100$ & $0.084$ & $0.078$ & $0.072$\\
\hline
$(10,5)$ & $0.4995$ & $0.3416$ & $0.109$ & $0.070$ & $0.056$ & $0.046$\\
\hline
$(14,7)$ & $0.5000$ & $0.2798$ & $0.109$ & $0.056$ & $0.039$ & $0.029$\\
\hline
$(6,5)$  & $0.8333$ & $0.5510$ & $0.264$ & $0.139$ & $0.102$ & $0.078$\\
\hline
\end{tabular}}
\caption{The statical and dynamical points for several regular codes 
and decoding algorithms, cf. Eqs. (\ref{BeliefPropagation1}), 
(\ref{BeliefPropagation2}).} 
\label{ThresholdsTab}
\end{table}
In Tab. \ref{ThresholdsTab} we report the threshold noise levels for several
regular codes, obtained using the density evolution method, together with
the thresholds for the optimal decoding strategy, see Ref.
\cite{GallagerAM}.

Finally let us notice that the fixed point of the iteration 
(\ref{DensityEvolution_1})-(\ref{DensityEvolution_2}) is the
replica symmetric order parameter for the equivalent spin model.
%
%*********************************************************************
%
\section{Statistical mechanics formulation and the replica approach}
\label{ReplicaSection}

We want to define a statistical mechanics model which
describes the decoding problem.
The probability distribution for the input codeword to be 
$\us = (\sigma_1,\dots,\sigma_N)$ conditional to the received message,
takes the form
\begin{eqnarray}
P(\us) = \frac{1}{Z}\delta_{\mathbb H}[\us] \, 
\exp\left\{\sum_{i=1}^{N}h_i\sigma_i\right\}\, ,
\label{ProbCodewords}
\end{eqnarray}
where $\delta_{\mathbb H}[\us]=1$ if $\us$ satisfies the parity checks 
encoded by the matrix ${\mathbb H}$, cf. Eq. (\ref{ParityCheckMatrix}),
and $\delta_{\mathbb H}[\us]=0$
otherwise. Since we assume the input codeword to be 
$\us^{\rm in} = (+1,+1,\dots,+1)$, the $h_i$ are i.i.d. with distribution
$p(h)$.

We modify the probability distribution (\ref{ProbCodewords}) in two ways:
\begin{enumerate}
\item We multiply the fields $h_i$ by a weight $\zh$. This allows 
us to tune the importance of the received message, analogously to Eqs. 
(\ref{BeliefPropagation1}) and (\ref{BeliefPropagation2}).
This modification was already considered in Ref. \cite{GallagerAM}.
Particularly important cases are $\zh=1$ and $\zh=0$.
\item We relax the constraints implied by the characteristic function 
$\delta_{\mathbb H}[\us]$. More precisely, let us denote each parity 
check by the un-ordered set of bits positions $(i_1,\dots,i_k)$
which appears in it. For instance the three parity checks in the Hamming code
${\cal H}_2(3)$, cf. Eq. (\ref{HammingMatrix}), are $(1,4,5,7)$,
$(2,4,6,7)$, $(3,5,6,7)$. Moreover let $\Omega_k$ be the set of all 
parity checks involving $k$ bits (in the irregular {\it ensemble}
the size of $\Omega_k$ is $Nc_k$). We can write explicitly the 
characteristic function $\delta_{\mathbb H}[\us]$ as follows:
\begin{eqnarray}
\delta_{\mathbb H}[\us] = \prod_{k=3}^{\infty} 
\prod_{(i_1\dots i_k)\in \Omega_k} 
\delta(\sigma_{i_1}\cdots\sigma_{i_k},+1)\, ,
\label{Hamiltonian}
\end{eqnarray}
where $\delta(\cdot,\cdot)$ is the Kronecker delta function.
Now it is very simple to relax the constraints by making
the substitution 
$\delta(\sigma_{i_1}\cdots\sigma_{i_k},+1)\to 
\exp\{\beta[\sigma_{i_1}\cdots\sigma_{i_k}-1]\}$.
\end{enumerate}
Summarizing the above considerations, we shall consider 
the statistical mechanics model defined by the Hamiltonian
\begin{eqnarray}
H(\sigma) = -\sum_{k=3}^{\infty}\sum_{(i_1\dots i_k)\in \Omega_k }
\!\!\!\! (\sigma_{i_1}\cdots\sigma_{i_k}-1)\;\;\;
-\;\frac{\zh}{\beta}\sum_{i=1}^N h_i\sigma_i\, ,
\label{Model}
\end{eqnarray}
at inverse temperature $\beta$.

We address this problem by the replica approach \cite{SpinGlass}
The replicated partition function reads
\begin{eqnarray}
\<Z^n\> \sim \int\!\prod_{\vs}d\l(\vs)d\lh(\vs)
\, e^{-N S[\l,\lh]}\, ,
\end{eqnarray}
with the action 
\begin{eqnarray}
S[\l,\lh] & = & \la\sum_{\vs}\l(\vs)\lh(\vs) - 
\frac{\la}{\ka}\sum_{k=3}^{\infty}c_k\sum_{\vs_1\dots\vs_k}
J_{\beta}(\vs_1,\dots,\vs_k)\l(\vs_1)\cdots\l(\vs_k)-\label{Action}\\
&&-\sum_{l=2}^{\infty} v_l\log\left[\sum_{\vs}\lh(\vs)^l {\cal H}(\vs)\right]
-\la+\frac{\la}{\ka}\, ,\nonumber
\end{eqnarray}
where
\begin{eqnarray}
J_{\beta}(\vs_1,\dots,\vs_k)  \equiv  
e^{\beta\sum_a(\sigma_1\dots\sigma_k-1)}\, ,\;\;\;\;\;\;\;\;
{\cal H}(\vs) = \<e^{\zh h\sum_a\sigma_a}\>_h\, ,
\end{eqnarray}
$\< \cdot \>_h$ being the average over $p(h)$.
The order parameters $\l(\vs)$ and $\lh(\vs)$ are closely related,
at least in the replica symmetric approximation, to the distribution
of messages in the decoding algorithm \cite{GallagerAM}, cf.
Eqs. (\ref{DensityEvolution_1}), (\ref{DensityEvolution_2}).
 
In the case of the BEC an irrelevant infinite constant must be
subtracted from the action (\ref{Action}) in order to get finite results.
This corresponds to taking
\begin{eqnarray}
{\cal H}_{BEC}(\vs)  \equiv  p + (1-p)\delta_{\vs,\vs_0}\, ,
\end{eqnarray}
where $\vs_0 = (+1,\dots,+1)$.
%
%*********************************************************************
%
\section{Binary erasure channel: analytical and numerical results}
\label{BECSection}

The binary erasure channel is simpler than the general case. 
Intuitively this happens because one cannot 
receive misleading indications concerning a bit. Nonetheless it 
is an important case both from the practical \cite{DigitalFountain}
and from the theoretical point of 
view \cite{FiniteLength,RichardsonUrbankeIntroduction,Tornado}.
%
%*************************
%
\subsection{The decoding algorithm}

Iterative decoding algorithms for irregular codes were 
first introduced and analyzed within this context 
\cite{Tornado}.
Belief propagation becomes particularly simple.
Since the knowledge about a received bit is completely sure,
the log-likelihoods $\{ h_i\}$, cf. Eq. (\ref{Likelihood}), 
take the values $h_i=+\infty$ 
(when the bit has been received\footnote{Recall that we are assuming the
channel input to be $\sigma^{\rm in}_i=+1$ for $i=1,\dots,N$.}) 
or $h_i=0$ (when it has been erased).
Analogously the messages $\{ x^{(t)}_{i\to\alpha}\}$ and 
$\{ y^{(t)}_{\alpha\to i}\}$ must assume the same two values. 
The rules (\ref{BeliefPropagation1}), (\ref{BeliefPropagation2}) become
\begin{eqnarray}
x^{(t+1)}_{i\to\alpha} & =&\left\{\begin{array}{rl}
+\infty & \mbox{ if either }h_i=+\infty\mbox{ or }
y^{(t)}_{\alpha'\to i}=+\infty
\mbox{ for some $\alpha'\ni i$ (with $\alpha '\neq\alpha$)},\\
0 & \mbox{otherwise,}\end{array}\right.\\
y^{(t+1)}_{\alpha\to i} & =&\left\{\begin{array}{rl}
+\infty & \mbox{ if }x^{(t)}_{j\to\alpha}=+\infty
\mbox{ for all the $j\in\alpha$ (with $j\neq i$)},\\
0 & \mbox{otherwise.}\end{array}\right.
\end{eqnarray}

There exists an alternative formulation \cite{Tornado} of the same algorithm.
Consider the system of $M$ linear equations (\ref{ParityCheckMatrix})
and eliminate from each equation the received variables (which are known 
for sure to be $\t0$). You will obtain a new linear system. In some 
cases you may have eliminated all the variables of one equation, the 
equation is satisfied and can therefore be eliminated.
For some of the other equations you may have eliminated all the 
variables but one. The remaining variable can be unambiguously fixed 
using this equation (since the received message is not misleading, this
choice is surely correct).  This allows to eliminate the variable from the 
entire linear system. This simple procedure is repeated until either all the
variables have been fixed, or one gets stuck on a linear system such that 
all the remaining equations involve at least two variables 
(this is called a {\it stopping set} \cite{FiniteLength}).

Let us for instance consider the linear system defined by the parity check
matrix (\ref{HammingMatrix}). Suppose, in a first case, that the received
message was $(\t0, \ast,\t0,\ast,\t0,\ast,\t0)$ (meaning that the bits of 
positions $2$, $4$, $6$ were erased). The decoding algorithm proceeds
as follows:
{\footnotesize
\begin{eqnarray}
\left\{\begin{array}{lcr}
\tx_1+\tx_4+\tx_5+\tx_7 & = & \t0\\
\tx_2+\tx_4+\tx_6+\tx_7 & = & \t0\\
\tx_3+\tx_5+\tx_6+\tx_7 & = & \t0
\end{array}
\right.\Rightarrow
\left\{\begin{array}{lcr}
\tx_4 & = & \t0\\
\tx_2+\tx_4+\tx_6 & = & \t0\\
\tx_6 & = & \t0
\end{array}
\right.\Rightarrow
\left\{\begin{array}{lcr}
\t0&=&\t0\\
\tx_2 & = & \t0\\
\t0 & = & \t0
\end{array}
\right.\, .
\end{eqnarray}}
In this case the algorithm succeeded in solving the decoding problem. 
Let us now see what happens if the received message is 
$(\ast,\t0,\ast,\t0,\ast,\t0,\ast)$:
{\footnotesize
\begin{eqnarray}
\left\{\begin{array}{lcr}
\tx_1+\tx_4+\tx_5+\tx_7 & = & \t0\\
\tx_2+\tx_4+\tx_6+\tx_7 & = & \t0\\
\tx_3+\tx_5+\tx_6+\tx_7 & = & \t0
\end{array}
\right.\Rightarrow
\left\{\begin{array}{lcr}
\tx_1+\tx_5+\tx_7 & = & \t0\\
\tx_7 & = & \t0\\
\tx_3+\tx_5+\tx_7 & = & \t0
\end{array}
\right.\Rightarrow
\left\{\begin{array}{lcr}
\tx_1+\tx_5&=&\t0\\
\t0 & = & \t0\\
\tx_3+\tx_5 & = & \t0
\end{array}
\right.\, .
\end{eqnarray}}
The algorithm found a stopping set. Notice that the
resulting linear system may well have a unique solution (although 
this is not the case in our example), which
can be found by means of simple polynomial algorithms \cite{NR}. 
Simply the iterative algorithm is unable to 
further reduce it.

\begin{figure}
\begin{tabular}{cc}
\epsfig{figure=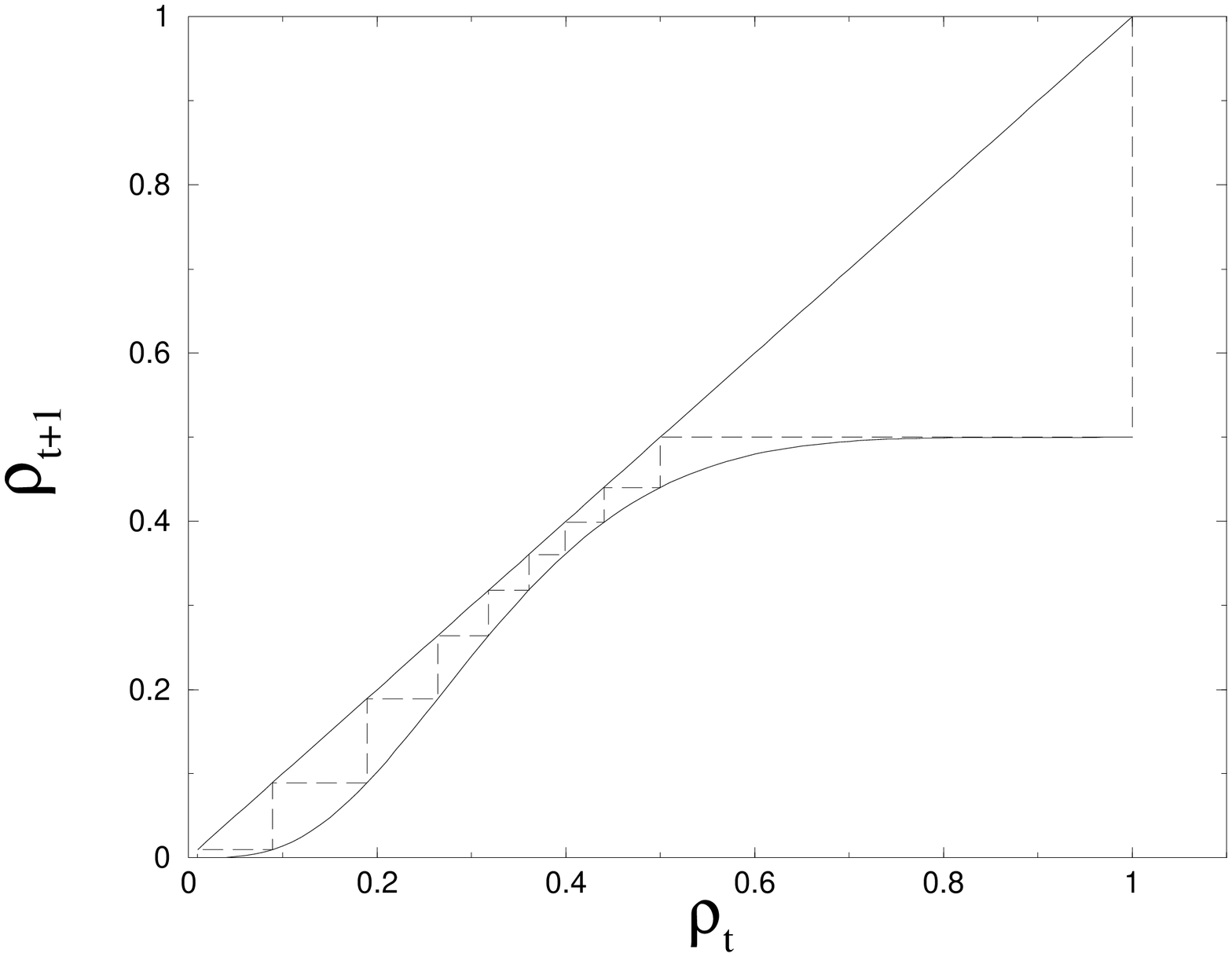,angle=-90,width=0.4\linewidth}&
\epsfig{figure=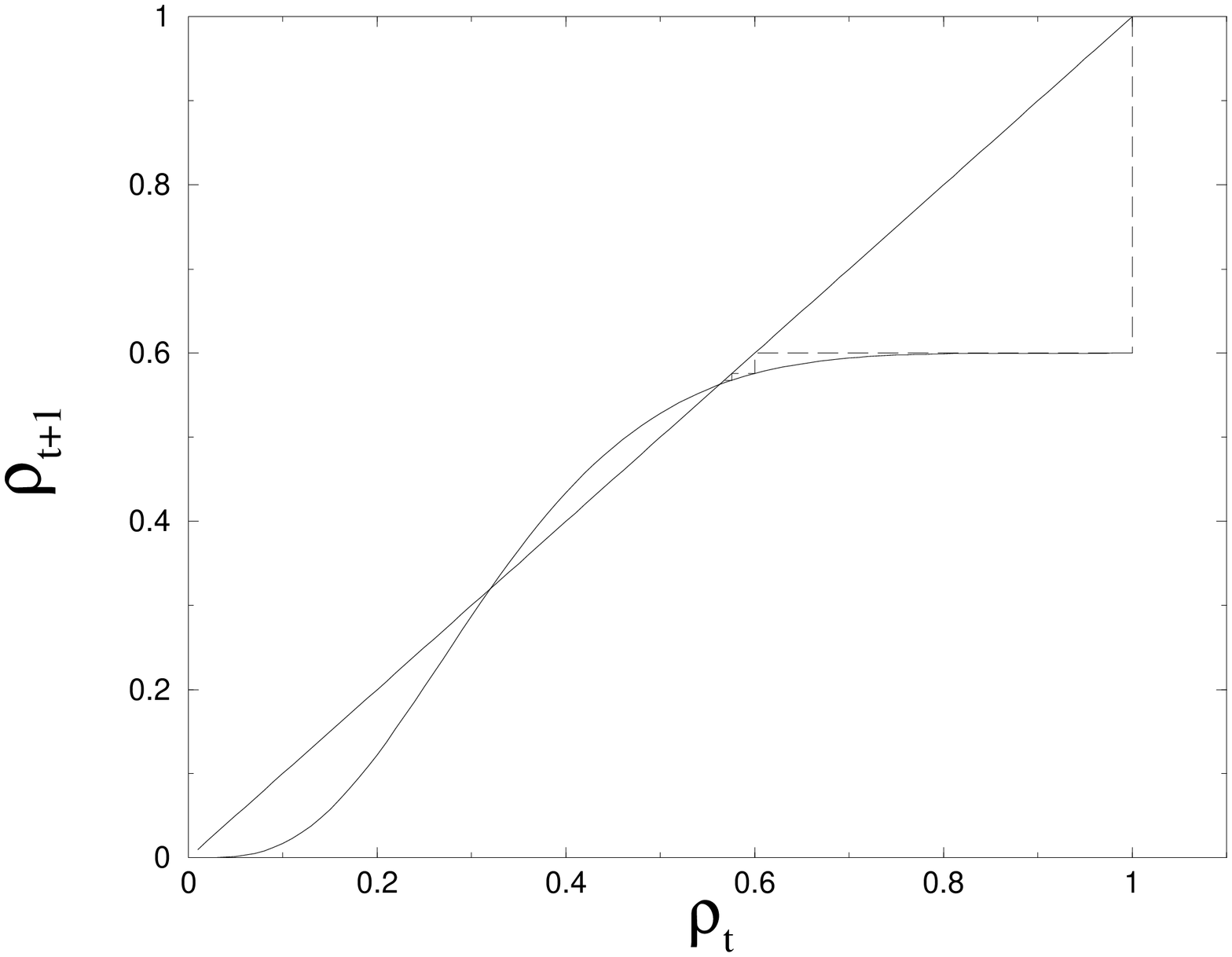,angle=-90,width=0.4\linewidth}
\end{tabular}
\caption{The evolution of the iterative decoding algorithm 
on the BEC, cf. Eqs. (\ref{DensityEvolutionBEC}). Here we
consider the $(6,5)$ code: $\rho_{t+1}=p[1-(1-\rho_t)^5]^4$.
On the left $p=0.5<p_d$, on the right $p=0.6>p_d$.}
\label{IterationBEC_Fig}
\end{figure}

The analysis of this algorithm \cite{RichardsonUrbankeIntroduction} uses the 
density evolution equations (\ref{DensityEvolution_1}), 
(\ref{DensityEvolution_2}) and is greatly simplified because the messages
$\{ x^{(t)}_{i\to\alpha}\}$ and 
$\{ y^{(t)}_{\alpha\to i}\}$ take only two values.
Their distributions have the form:
\begin{eqnarray}
\pi_t(x) = \rho_t\,\delta(x)+(1-\rho_t)\,\delta_{\infty}(x)\;\;\; ,
\ph_t(x) = \rh_t\,\delta(y)+(1-\rh_t)\,\delta_{\infty}(y)\, ,
\end{eqnarray}
where $\delta_{\infty}(\cdot)$ is a delta function centered at $+\infty$.
The parameters $\rho$ and $\rh$ give the fraction of zero messages,
respectively from variables to checks and from checks to variables.
Using Eqs. (\ref{DensityEvolution_1}) and (\ref{DensityEvolution_2}),
we get:
\begin{eqnarray}
\rho_{t+1} = p\,\frac{v'(\rh_t)}{v'(1)}\;\, ,\;\;\;\;\;\;\;\;\;
\rh_{t+1} = 1-\frac{c'(1-\rho_t)}{c'(1)}\, .
\label{DensityEvolutionBEC}
\end{eqnarray}
The initial condition $\rho_0=\rh_0=1$ converges to the perfect
recovery fixed point $\rho=\rh=0$ if $p<p_d$. This corresponds to 
perfect decoding. For $p>p_d$ the algorithm gets stuck
on a non-trivial linear system:  $\rho_t\to\rho_*$, 
$\rh_t\to\rh_*$, with $0<\rho_*,\rh_*<1$. The two regimes 
are illustrated in Fig. \ref{IterationBEC_Fig}.
%
%
%*************************
%
\subsection{Statical transition}

In the spin model corresponding to the situation described above,
we have two types of spins: the ones corresponding to correctly 
received bits, which are fixed by an infinite magnetic field
$h_i=+\infty$; and the ones corresponding to erased bits, on which 
no magnetic field acts: $h_i=0$. We can therefore consider 
an effective model for the erased bits once the received ones are 
fixed to $+1$. This correspond somehow to what is done by the decoding 
algorithm: the received bits are set to their values in the very
first step of the algorithm and remain unchanged thereafter.

Let us consider the zero temperature limit.
If the system is in equilibrium, 
its probability distribution will concentrate on zero energy 
configurations: the codewords. We will have typically 
${\cal N}_{\rm words}(p)\sim 2^{Ns_{\rm words}(p)}$ 
codewords compatible with the received 
message. Their entropy $s_{\rm words}(p)$ can be computed within the 
replica formalism, cf. App. \ref{BECAppendix}. The result is
\begin{eqnarray}
s_{\rm words}(\rho,\rh;p) =
\la\rho(1-\rh)+\frac{\la}{\ka}\,c(1-\rho)+p\,v(\rh)-\frac{\la}{\ka}\, ,
\label{ParaEntropy}
\end{eqnarray}
which has to be maximized with respect to the order parameters
$\rho$ and $\rh$. The saddle point equations have exactly 
the same form as the fixed point equations corresponding to the
dynamics (\ref{DensityEvolutionBEC}), namely 
$\rho = p v'(\rh)/v'(1)$ and $\rh = 1-c'(1-\rho)/c'(1)$

The saddle point equations have two stable solutions, i.e. local maxima of
the entropy (\ref{ParaEntropy}): 
$(i)$ a completely ordered solution 
$\rho =\rh=0$, with entropy $s_{\rm words}(0,0)=0$
(in some cases this solution becomes locally unstable above some noise 
$p_{loc}$); $(ii)$ (for sufficiently
high noise level) a paramagnetic solution $\rho_*,\rh_*>0$. 
The paramagnetic solution appears at the same value $p_d$ of the noise
above which the decoding algorithm gets stuck. 

The fixed point 
to which the dynamics (\ref{DensityEvolutionBEC}) converges
coincides with the statistical mechanics result for $\rho_*,\rh_*$.
However the entropy of the paramagnetic solution 
$s_{\rm words}(\rho_*,\rh_*)$ is negative at $p_d$ and becomes
positive only above a certain critical noise $p_c$. This means that
the linear system produced by the algorithm continues to have a 
unique solution below $p_c$, although our linear time algorithm is unable
find such a solution.

The ``dynamical'' critical noise $p_d$ is the solution of the
following equation
\begin{eqnarray}
p\frac{v''(\rh_*)c''(1-\rho_*)}{v'(1)c'(1)} = -1\, ,
\end{eqnarray}
where $\rho_*$ and $\rh_*$ solve the saddle point equations.
The statical noise can be obtained setting $s_{\rm words}(\rho_*,\rh_*)=0$.
Finally the completely ordered solution becomes locally unstable
for
\begin{eqnarray}
p_{loc} = \frac{c'(1)v'(1)}{v''(0)c''(1)}\, .
\end{eqnarray}
\begin{figure}
\centerline{\hspace{-2cm}
\epsfig{figure=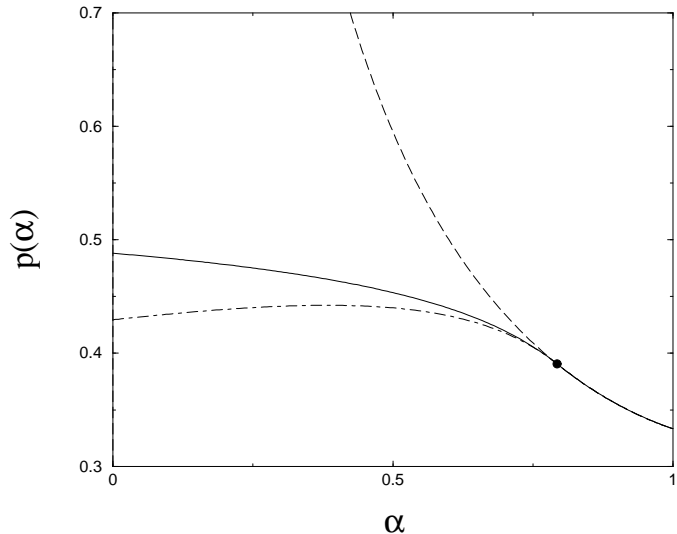,angle=-90,width=0.5\linewidth}}
\caption{The phase diagram of the family of codes with generating
polynomials $c(x) = \alpha x^4+(1-\alpha)x^6$,
$v(x) = \alpha x^2+(1-\alpha)x^3$.
The dashed line gives the local stability threshold for the
completely ordered ferromagnetic phase. The continuous and dot-dashed
lines refer (respectively) to the static and dynamic critical points
$p_c(\alpha)$ and $p_d(\alpha)$.}
\label{Critical}
\end{figure}
As an example let us consider the one-parameter family of
$R = 1/2$ codes specified by the
following generating polynomials: $c(x) = \alpha x^4+(1-\alpha)x^6$,
$v(x) = \alpha x^2+(1-\alpha)x^3$. This is an irregular code which smoothly
interpolates between the regular $(6,3)$ and $(4,2)$ codes.
The local stability threshold is given by
\begin{eqnarray}
p_{loc}(\alpha) = \frac{(3-\alpha)^2}{6\alpha(5-3\alpha)}\, .
\end{eqnarray}
The dynamical and critical curves $p_d(\alpha)$ and $p_c(\alpha)$ are
reported in Fig. \ref{Critical}.
Notice that the $\alpha$ value where $p_d(\alpha)$ reaches its
maximum, corresponding to the best code in this family, is neither 0
nor 1.  This is a simple example showing that irregular codes ($0 <
\alpha < 1$) are generally superior to regular ones ($\alpha=0$ or
$\alpha=1$ in this example).
Notice also that above the tricritical point $\alpha_t \approx 0.79301412$,
$p_t \approx 0.39057724$ the three curves $p_{loc}(\alpha)$,
$p_c(\alpha)$ and $p_d(\alpha)$ coincide.
In the following we shall study in some detail the $\alpha = 0$
case, which corresponds to a regular $(6,3)$ code, the corresponding critical 
and dynamical points $p_c$ and $p_d$ are given in Tab. \ref{ThresholdsTab}. 
%
%*************************
%
\subsection{Dynamical transition}

The dynamical transition is not properly described within the 
replica symmetric treatment given above. Indeed, the paramagnetic 
solution cannot be considered, between $p_d$ and $p_c$, as a metastable state
because it has negative entropy. One cannot therefore give a 
sensible interpretation of the coincidence between the critical 
noise for the decoding algorithm, and the appearance of the 
paramagnetic solution.

Before embarking in the one step replica symmetry-breaking (1RSB)
calculation, let us review some well-known 
facts \cite{MonassonMarginal,FranzParisi}.
Let us call $m\phi(\beta,m)$ the free energy of 
$m$ weakly coupled ``real'' replicas times beta. 
This quantity can be computed in 1RSB calculation. 
In the limit $\beta\to\infty$, with $m\beta=\mu$ fixed,
we have $m\phi(\beta,m)\to\mu\phi(\mu)$.
The number of metastable states with a given energy density
$\epsilon$ is
\begin{eqnarray}
{\cal N}_{MS}(\epsilon) \sim e^{N\Sigma(\epsilon)}\, ,
\end{eqnarray}
where the complexity $\Sigma(\epsilon)$ is the Legendre transform of
the $m$ replicas free energy:
\begin{eqnarray}
\Sigma(\epsilon) = \left.\mu\epsilon-
\mu\phi(\mu)\right|_{\epsilon = \partial[\mu\phi(\mu)]}\,
.\label{Complexity}
\end{eqnarray}
The (zero temperature) dynamic energy $\epsilon_d$ and the static 
energy $\epsilon_s$
are\footnote{Notice that one can give (at least) three possible definitions
of the dynamic energy: $(i)$ from the solution of the nonequilibrium 
dynamics: $\epsilon_d^{(d)}$; $(ii)$ imposing the replicon eigenvalue to
vanish: $\epsilon^{(r)}_d$; $(iii)$ using, as in the text,
the complexity $\Sigma(\epsilon)$: $\epsilon^{(c)}_d$.
The three results coincide in the $p$-spin spherical fully connected 
model, however their equality in the present case is, at most, a conjecture.},
respectively, the maximum and the minimum energy such that 
$\Sigma(\epsilon)\ge 0$.

The static energy is obtained by solving the following equations:
\begin{eqnarray}
\left\{\begin{array}{c}
\epsilon_s = \phi(\mu)\, ,\\
\partial\phi(\mu)= 0\, ,
\end{array}\right.
\end{eqnarray}
which corresponds to the usual prescription of maximizing the free
energy over the replica symmetry breaking parameter $m$ \cite{SpinGlass}. 
The dynamic energy is given by
\begin{eqnarray}
\left\{\begin{array}{c}
\epsilon_d = \partial[\mu\phi(\mu)]\, ,\\
\partial^2[\mu\phi(\mu)] = 0\, .
\end{array}\right.
\end{eqnarray}
Finally, if $\epsilon_s=0$ the complexity of the ground state is
$\Sigma(0)=-\lim_{\mu\to\infty}\mu\phi(\mu)$.

We weren't able to exactly compute the 1RSB free energy $\phi(\mu)$.
However excellent results can be obtained within an ``almost factorized''
variational Ansatz, cf. App. \ref{BEC_RSB}. 
The picture which emerges is the following:
\begin{itemize}
\item In the low noise region ($p<p_d$), no metastable states exist.
Local search algorithms should therefore be able to recover the
erased bits.
\item In the intermediate noise region ($p_d<p<p_c$) an 
exponentially large number of metastable states appears. They 
have energy densities $\epsilon$ in the range 
$\epsilon_s<\epsilon<\epsilon_d$, with $\epsilon_s>0$. Therefore the 
transmitted codeword is still the only one compatible with the received 
message. Nonetheless a large number of extremely stable {\it pseudo-codewords}
stop local algorithms. The number of violated parity checks 
in these codewords cannot be reduced by means of local moves.
\item  Above $p_c$ we have $\epsilon_s=0$: a fraction of the metastable
states is made of codewords. Moreover $\Sigma(0)$ (which gives the number of
such codewords) coincides with the paramagnetic
entropy $s_{\rm words}(\rho_*,\rh_*)$ computed in the previous Section.
\end{itemize}
\begin{figure}
\centerline{\hspace{-2cm}
\epsfig{figure=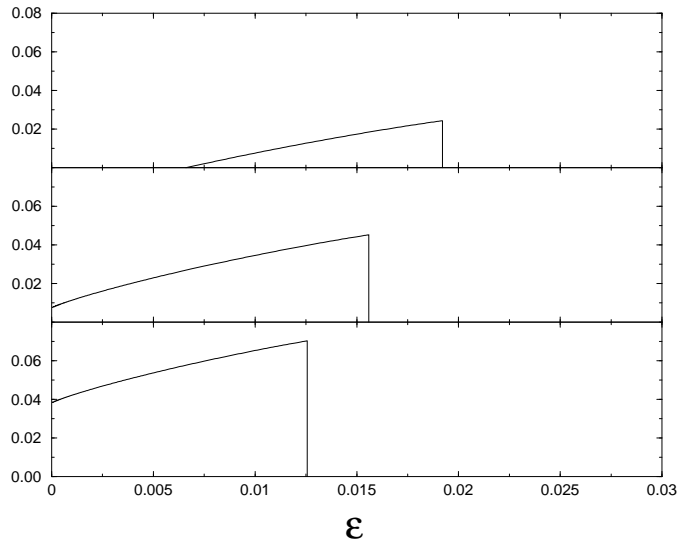,angle=-90,width=0.5\linewidth}}
\caption{The complexity $\Sigma(\epsilon)$ for (from top to bottom) 
$p=0.45$ (below $p_c$), $p = 0.5$, and $p=0.55$ (above $p_c$).}
\label{ComplexityVSEnergy}
\end{figure}
As an illustration, let us consider the $(6,3)$ regular code.
In Fig. \ref{ComplexityVSEnergy} we plot the resulting complexity
curves $\Sigma(\epsilon)$ for three different values of the erasure
probability $p$.
\begin{figure}
\begin{tabular}{cc}
\epsfig{figure=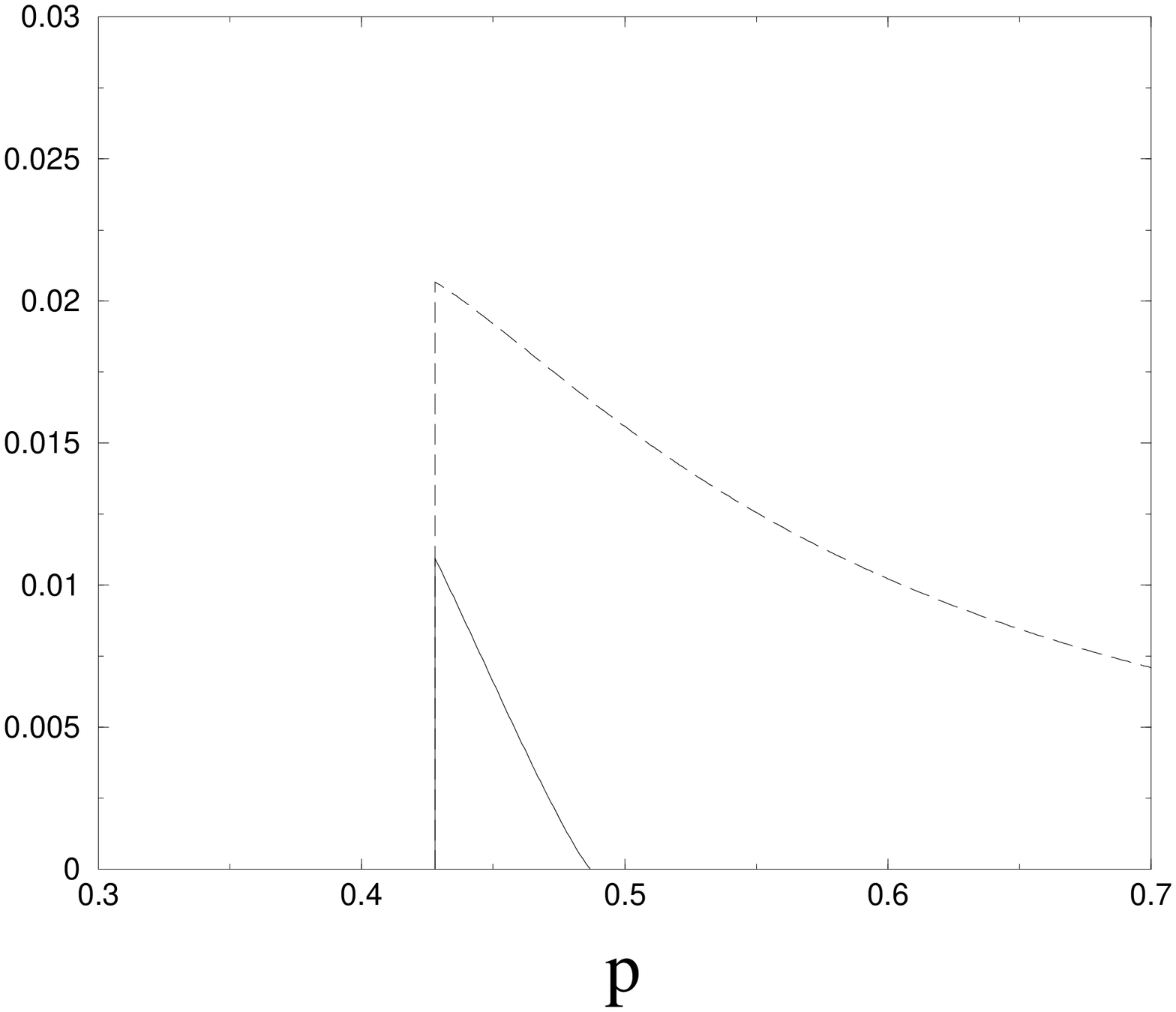,angle=-90,width=0.45\linewidth}&
\epsfig{figure=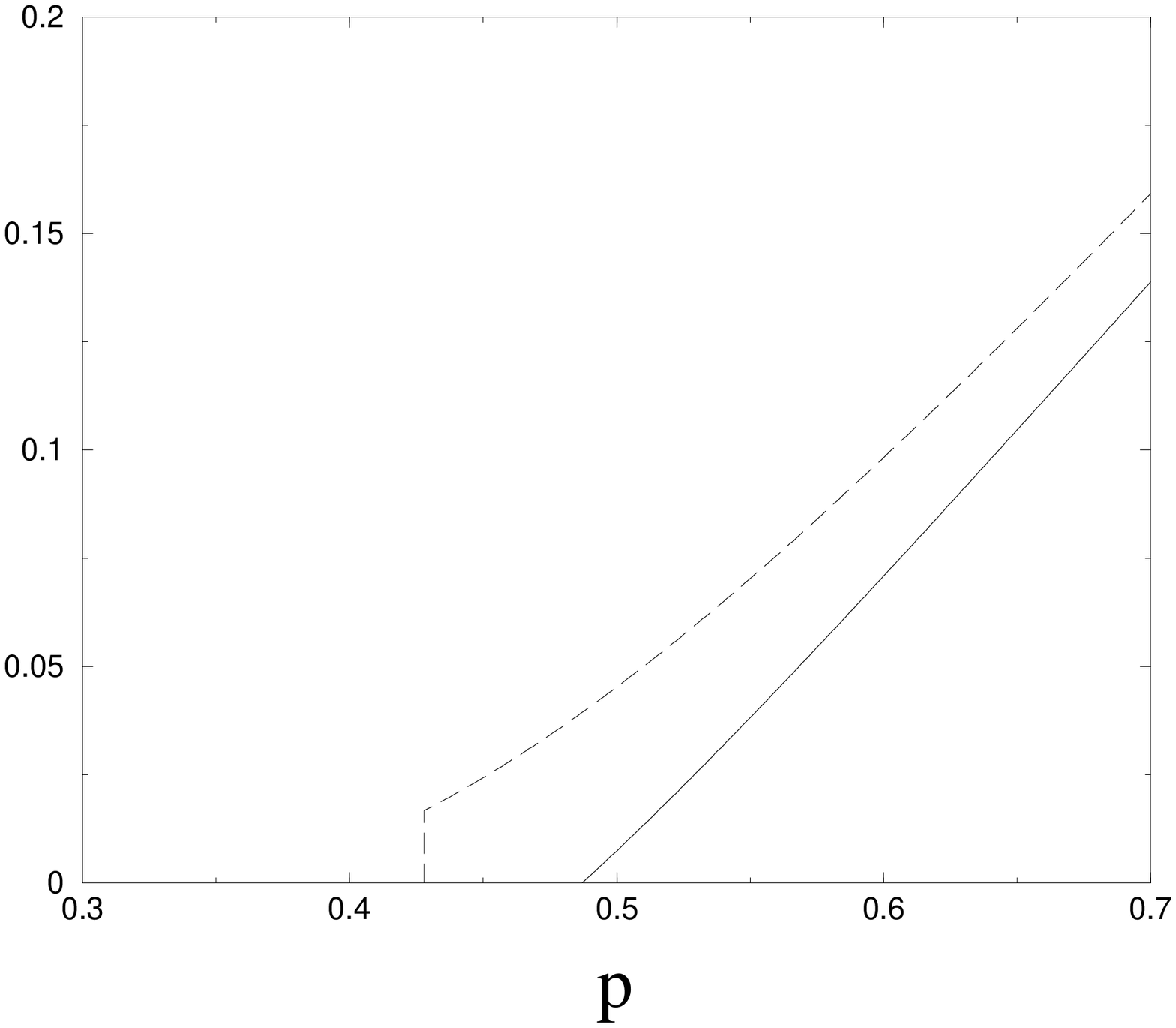,angle=-90,width=0.45\linewidth}
\end{tabular}
\caption{Left-hand frame:
the static and dynamic energies $\epsilon_s$ and $\epsilon_d$
of the metastable states (respectively, solid and dashed lines).
Right-hand frame: the total complexity $\max_{\epsilon}
\Sigma(\epsilon)$ and the zero energy complexity $\Sigma(0)$.}
\label{DynamicE_StaticE}
\end{figure}
In Fig. \ref{DynamicE_StaticE}, left frame, we report the static and dynamic
energies $\epsilon_s$ and $\epsilon_d$ as functions of $p$.
In the right frame we present the total complexity $\Sigma_{{\rm tot}}\equiv
\max_{\epsilon} \Sigma(\epsilon)=\Sigma(\epsilon_d)$, and the zero
energy complexity $\Sigma(0)$.
%
%****************************
%
\subsection{Numerical results}
\label{NumericalBEC}

In order to check analytical predictions and to better illustrate the
role of metastable states, we have run a set of Monte Carlo
simulations, with Metropolis dynamics, on the Hamiltonian (\ref{Model}) of
the (6,3) regular code for the BEC. Notice that local search algorithms
for the decoding problem have been already considered by the 
coding theory community \cite{ExpanderCodes}.

We studied quite large codes ($N=10^4$ bits), 
and tried to decode it (i.e. to find a ground state of the corresponding 
spin model) with the help
of simulated annealing techniques~\cite{SIM_ANN}.  For each value of
$p$, we start the simulation fixing a fraction $(1-p)$ of spins to
$\sigma_i = +1$ (this part will be kept fixed all along the run).  The
remaining $p N$ spins are the dynamical variables we change during the
annealing in order to try to satisfy all the parity checks.  The
energy of the system counts the number of unsatisfied parity checks.

The cooling schedule has been chosen in the following way: $\tau$
Monte Carlo sweeps (MCS)~\footnote{Each Monte Carlo sweep consists in
$N$ proposed spin flips. Each proposed spin flip is accepted or not 
accordingly to a standard Metropolis test.} 
at each of the 1000 equidistant temperatures
between $T=1$ and $T=0$.  The highest temperature is such that the
system very rapidly equilibrates on the paramagnetic energy
$\epsilon_P(T)$.  Typical values for $\tau$ are from 1 to $10^3$.

Notice that, for any fixed cooling schedule, the computational complexity of
the simulated annealing method is linear in $N$.  Then we expect it to be
affected by metastable states of energy $\epsilon_d$, which are
present for $p>p_d$: the energy relaxation should be strongly reduced
around $\epsilon_d$ and eventually be completely blocked.

\begin{figure}
\begin{tabular}{cc}
\hspace{-1.5cm}\epsfig{figure=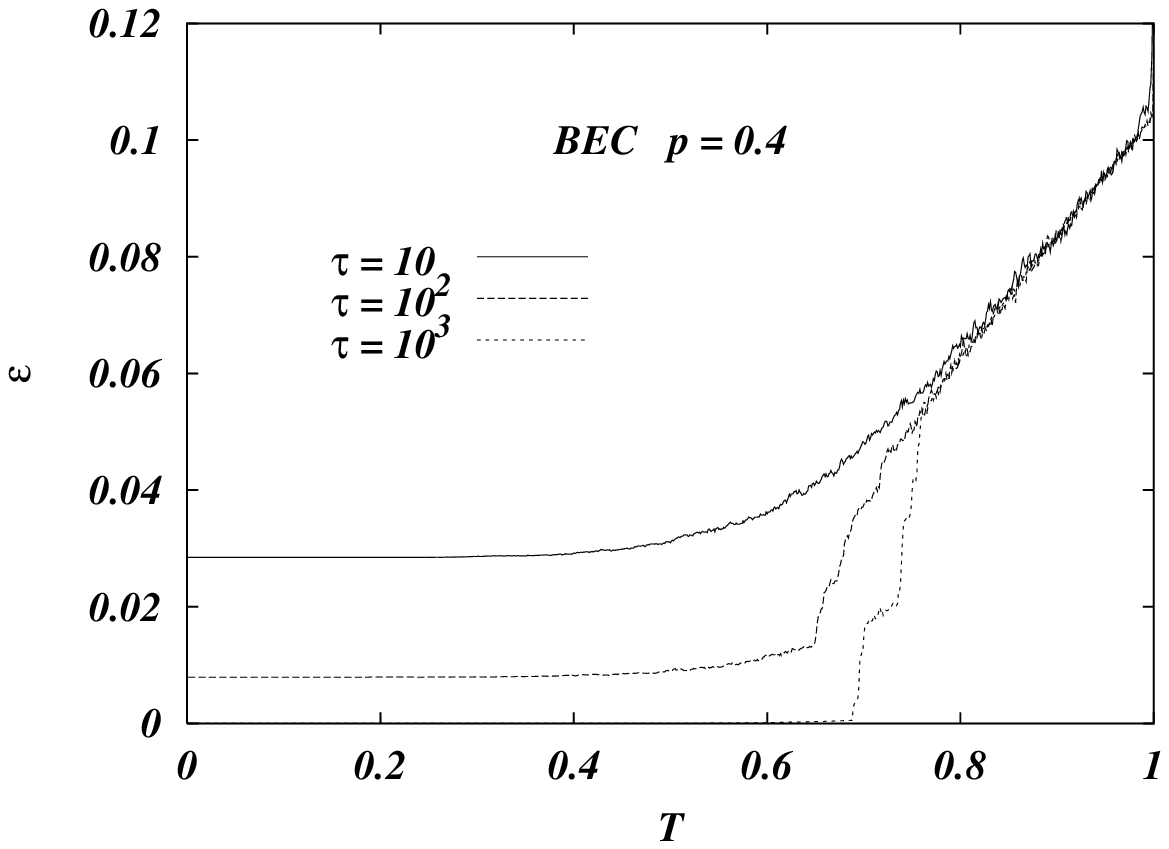,angle=0,width=0.5\linewidth}&
\hspace{-0.5cm}\epsfig{figure=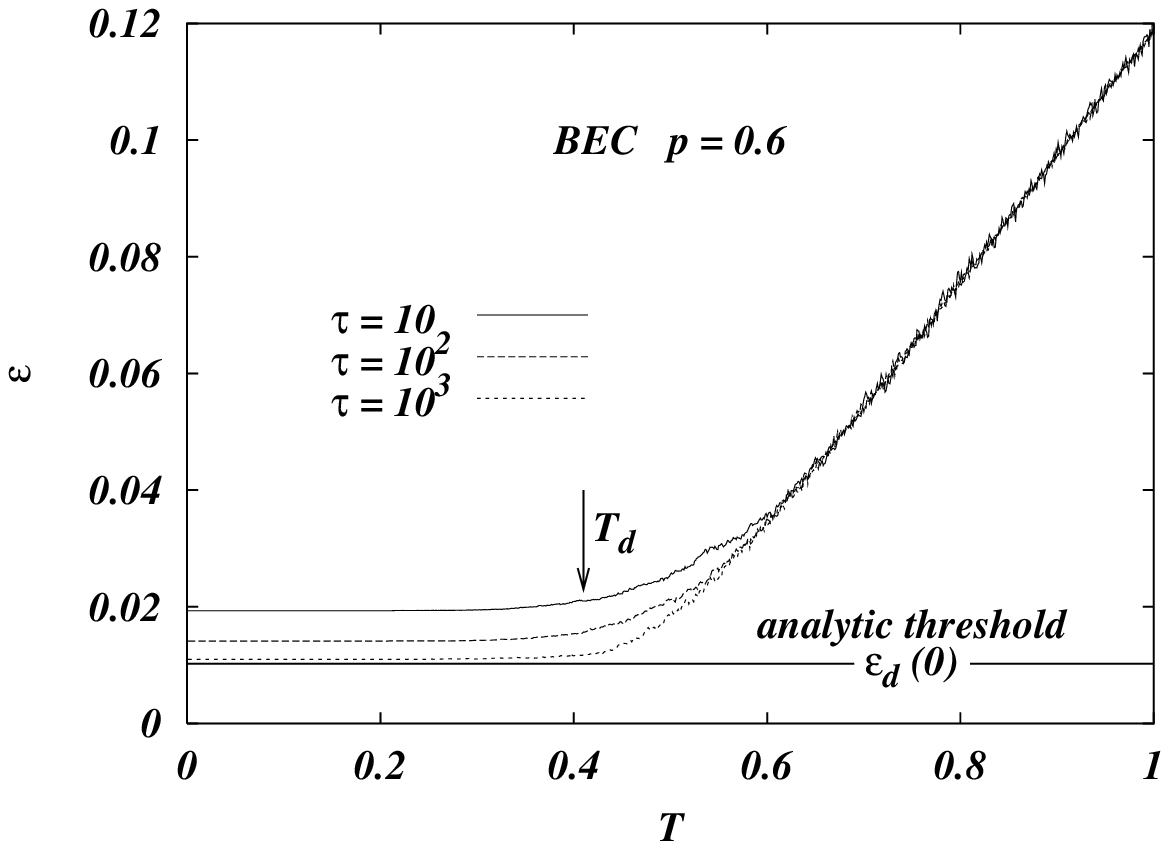,angle=0,width=0.5\linewidth}
\end{tabular}
\caption{Energy relaxation for the Hamiltonian of the (6,3) regular
code during the simulated annealing with $\tau$ MCS per temperature
and 1000 equidistant temperatures in $[0,1]$}
\label{bec_ann}
\end{figure}
In order to illustrate how the system relaxes during the simulated
annealing we show in Fig.~\ref{bec_ann} the energy density as a
function of the temperature for $p=0.4$ (left) and $p=0.6$ (right) and
various cooling rates, $\tau=10, 10^2, 10^3$ (each data set is the
average over many different samples).

For $p=0.4<p_d$ the final energy strongly depends on the cooling rate
and the slowest cooling procedure is always able to bring the system
on the ground state, corresponding to the transmitted codeword.
Decoding by simulated annealing is therefore successful.

For $p=0.6>p_d$ the situation drastically changes.  Below a
temperature $T_d$ (marked by an arrow in Fig.~\ref{bec_ann}, right frame)
there is an almost complete stop of the energy relaxation.  $T_d$
marks the dynamical transition and the corresponding energy
$\epsilon_d(T_d) = \epsilon_P(T_d)$ is called the threshold energy.
The energy of threshold states still varies a little bit with
temperature, $\epsilon_d(T)$, and the final value reached by 
the simulated annealing algorithm is its zero-temperature limit 
$\epsilon_d(0) = \epsilon_d$.
Remember that, by construction, ground states of zero energy are
present for any $p$ value, but they become unreachable for $p>p_d$,
because they become shielded by metastable states of higher energy.

\begin{figure}
\centerline{\epsfig{figure=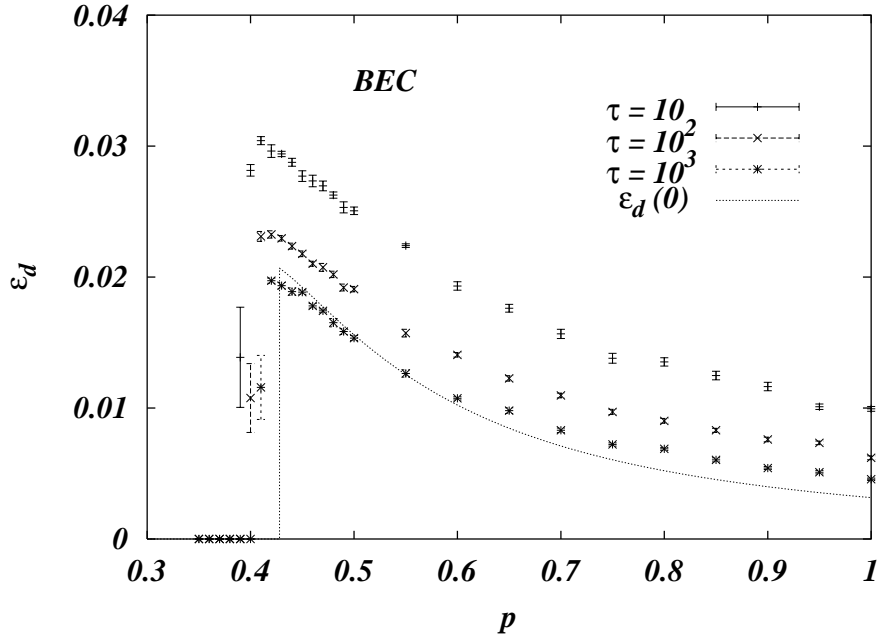,angle=0,width=0.7\linewidth}}
\caption{Lowest energies reached by the simulated annealing. Errors
are sample to sample fluctuations.}
\label{bec_e_min}
\end{figure}

We show in Fig.~\ref{bec_e_min} the lowest energy reached by the
simulated annealing procedure for different $p$ and $\tau$ values.
While for $p<p_d$ all parity checks can be satisfied and the energy
relaxes to zero in the limit of a very slow cooling, for $p \ge p_d$
the simulation get stuck in a metastable state of finite energy, that
is with a number of unsatisfied parity checks of order $N$.  The
agreement with the analytic prediction (dotted line) is quite good
everywhere, but very close to $p_d$.

Discrepancies between analytical predictions and numerical results may
be very well due to finite-size effects in the latter.  One possible
explanation for large finite-size effects near the dynamic critical point 
$p_d$ is the following.
Metastable states of energy $\epsilon_d$ are stable under any local dynamic,
which may flip simultaneously only a finite number of spins, and under
global dynamics flipping no more than $\omega N$ spins simultaneously.
Physical intuition (threshold states become more robust increasing
$p$) imply that the function $\omega(p)$ must monotonously increase
for $p \in [p_d,1]$.  Moreover, continuity reasons tell us that
$\omega(p_d)=0$.  The fact that $\omega(p)$ is very small close to
$p_d$, together with the fact that in numerical simulations we are
restricted to finite values of $N$, allow the local Monte Carlo
dynamic to relax below the analytical predicted threshold energy.  A
more detailed characterization of this effect is presently under study
and will be presented in a forthcoming publication.
%
%*********************************************************************
%
\section{The general channel: analytical and numerical results}
\label{GeneralChannelSection}

We considered the case of a general noisy channel
using two different approaches:
a finite-temperature and a zero-temperature approach.
While the first one offers a clear connection with the dynamics 
of decoding-by-annealing algorithm, the second one gives a nice 
geometrical picture of the situation.
%
%****************************
%
\subsection{Finite temperature}

Suppose you received some message encoded using a Gallager code and you 
want to decode it, but no one explained to you the belief propagation 
algorithm, cf. Eqs. (\ref{BeliefPropagation1}), 
(\ref{BeliefPropagation2}). 

A physicist idea would be the following.
Write the corresponding Hamiltonian $H(\us)$, see Eq. (\ref{Model}), 
and run a Monte Carlo algorithm at inverse temperature $\beta$.
If you wait enough time, you will be able to sample 
the configuration $\us$ according to the Boltzmann distribution
$P_{\beta}(\us) \propto e^{-\beta H(\us)}$. Then cool down the system
adiabatically: i.e. change the temperature according to some schedule
$\{\beta_1,\beta_2,\dots,\}$ with $\beta_k\uparrow\infty$,
waiting enough time at each temperature for the system to equilibrate.

As $\beta\to\infty$ the Boltzmann measure of the Hamiltonian 
(\ref{Hamiltonian}) concentrates on the codewords (for which the
exchange term in Eq. (\ref{Hamiltonian}) is equal to zero).
Moreover each codeword is given a weight which depends on its likelihood.
In formulae:
\begin{eqnarray} 
\lim_{\beta\to\infty}P_{\beta}(\us) = 
\frac{1}{Z_{\zh}}\, P(\us|\utx^{\rm out})^{\zh}\, ,
\end{eqnarray}
where $P(\us|\utx^{\rm out})$ is the probability for $\us$ to be the
transmitted codeword, conditional to the received message $\utx^{\rm out}$,
and $Z_{\zh}$ is a normalization constant.
Therefore when $\beta\gg 1$, our algorithm will sample a 
codeword with probability proportional to $P(\us|\utx^{\rm out})^{\zh}$.
For good codes below the critical noise threshold $p_c$, the
likelihood $P(\us|\utx^{\rm out})$ is strongly 
concentrated\footnote{Namely we have $P(\us^{\rm in}|\utx^{\rm out})
=1-O(e^{-\alpha N})$. This happens because there is
a minimum $O(N)$ Hamming distance between distinct codewords 
\cite{GallagerThesis}.} 
on the correct input codeword. 
Therefore the system will spend most of its time on the correct codeword
as soon as $\beta\gg 1$ and $\zh\ge 1$ (for $\zh<1$, $p_c$ has a non-trivial dependence on $\zh$, cf. Ref. \cite{GallagerAM}).

This algorithm will succeed as long as we are able to keep the system 
in equilibrium at all temperatures down to zero. If some form of ergodicity 
breaking is present this may take an exponentially (in the size $N$) 
long time.
Let us suppose to spend an $O(N)$ computational time at each temperature 
$\beta_i$ of the annealing schedule (this is what happens in Nature).
We expect to be able to equilibrate the system only
at low enough noise (let us say for $p<p_d(\zh)$), 
when the magnetic field in Eq. (\ref{Model})
is strong enough for single out a unique ergodic component.

\subsubsection{The random linear code limit}
\label{RLCbetaSection}

Some intuition on the static phase diagram can be gained
by looking at the $k,l\to\infty$ limit with rate $R=1-l/k$ fixed, cf. 
App. \ref{RLC_FiniteTemp_App}. 
Unhappily, in this limit the dynamic phase transition disappears:
the decoding algorithm is always unsuccessful, as can be understood
by looking at Eqs. (\ref{BeliefPropagation1})-(\ref{BeliefPropagation2}).
This phenomenon is analogous to what happens in the random energy model (REM)
\cite{DerridaREM}:
the dynamic transition is usually said to occur at infinite 
temperature. We refer to Sec. \ref{ZeroTRLC} for further clarifications 
of this point.

There exist a paramagnetic and a ferromagnetic phases, with free energy
densities
\begin{eqnarray}
f_P & = & -\frac{1}{\beta}\<\log (2\cosh \zh h)\>_h+\frac{1-R}{\beta}
\log(1+\tanh\beta)\, ,\label{FreeParaRLC}\\
f_F & = & -\frac{\zh}{\beta}\< h\>_h\, .\label{FreeFerroRLC}
\end{eqnarray}
One must be careful in computing the entropy because of the explicit
dependence of the Hamiltonian (\ref{Hamiltonian}) upon the temperature.
The result is that the ferromagnetic phase has zero entropy
$s_F = 0$, while the entropy of the paramagnetic phase is
\begin{eqnarray}
s_P & = & \<\log(2\cosh \zh h)\>_h-\<\zh h\tanh\zh h\>_h-\\
&& - (1-R)\log(1+\tanh\beta)+(1-R)\beta(1-\tanh\beta)\, .\nonumber
\end{eqnarray}
\begin{figure}
\begin{tabular}{cc}
\epsfig{figure=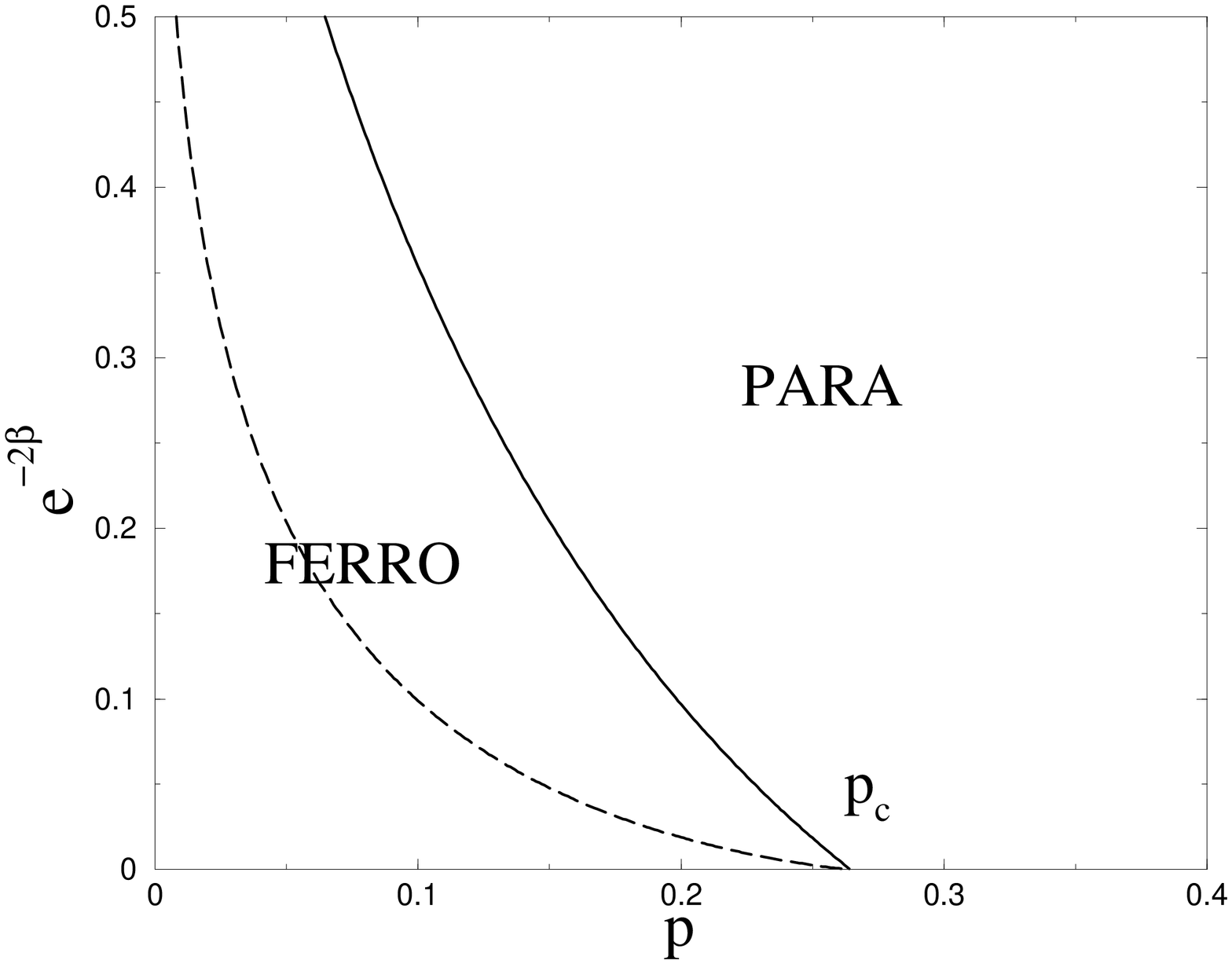,angle=-90,width=0.45\linewidth}&
\epsfig{figure=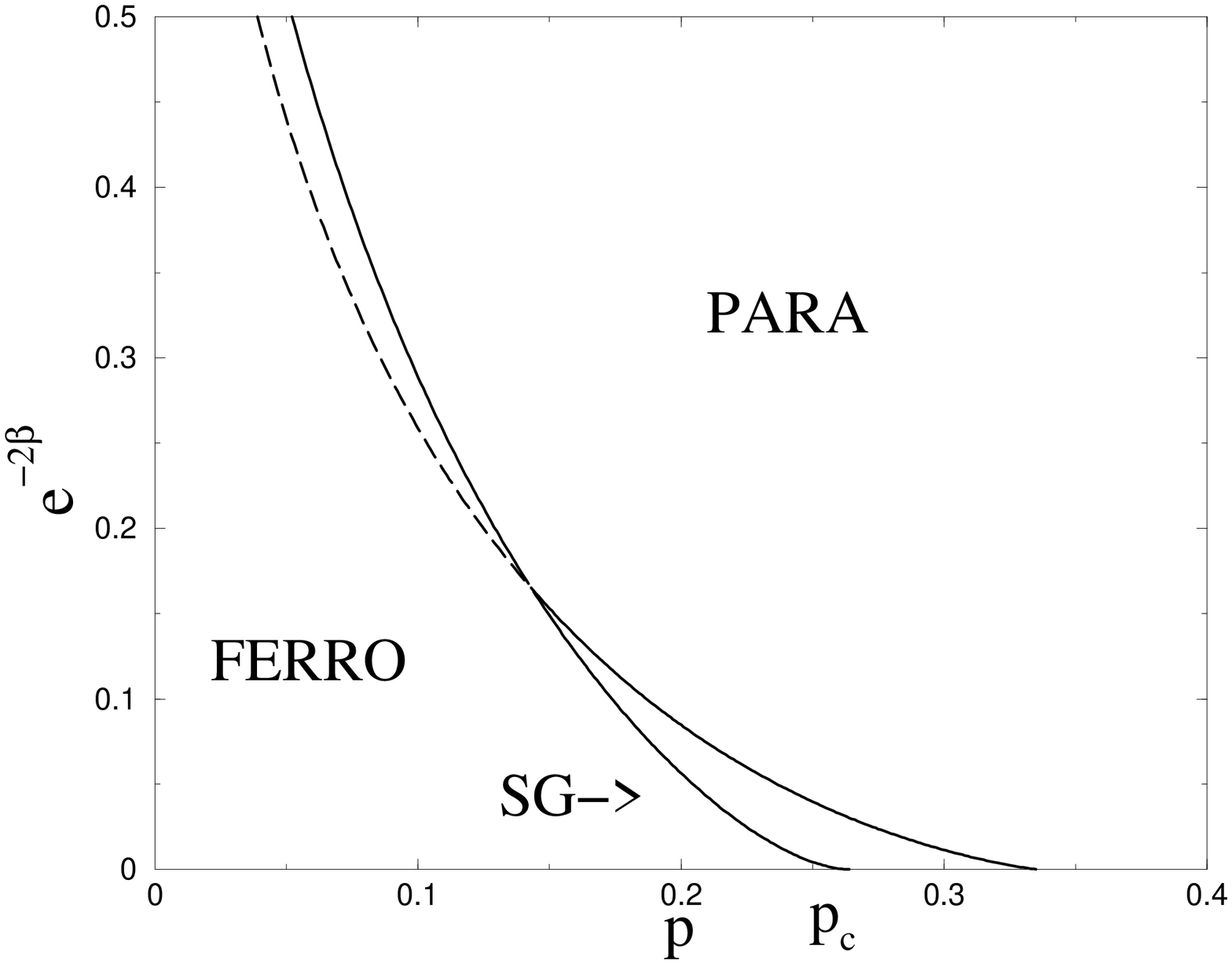,angle=-90,width=0.45\linewidth}
\end{tabular}
\caption{The phase diagram for the model (\ref{Hamiltonian}) in the limit
$k,l\to\infty$ with $R=1-l/k$ fixed. Here we consider $R=1/6$ and 
$\zh=1$ (on the left) and $1.5$ (on the right). The rightmost (i.e. noisier)
point for which the ferromagnetic phase is globally stable is always
at $\beta=\infty$, $p=\delta_{GV}(R)\approx 0.264$. Along the dashed line the
entropy of the paramagnetic phase vanishes.}
\label{PhaseDiagramRCL}
\end{figure}
In the low-temperature, low-noise region the paramagnetic entropy 
$s_P$ becomes negative. This signals a REM-like glassy transition
\cite{DerridaREM}. 
The spin glass free energy is obtained by maximizing over 
the RSB parameter $m$ (with $0\le m\le 1$)
the following expression
\begin{eqnarray}
f_{SG}(m) = -\frac{(1-R)}{\beta m}\log(1+e^{-2\beta m})
-\frac{1}{m}\<\log(2\cosh m\zh h )\>_h\, .\label{FreeSGRLC}
\end{eqnarray}

The generic phase diagram is reported in Fig. \ref{PhaseDiagramRCL}.
At high temperature, as the noise level is lowered the system undergoes
a paramagnetic-ferromagnetic transition and concentrates on the correct codeword.
At low temperature an intermediate glassy phase may be present (for $\zh>1$):
the system concentrates on a few incorrect configurations. 

\subsubsection{Theoretical dynamical line}
\label{CompleteFiniteBeta}

The existence of metastable states can be detected within 
the replica formalism by the so-called marginal stability condition.
One considers the saddle point equations for the 1RSB order parameter, 
fixing the RSB parameter $m=1$, cf. App. \ref{GeneralAppendix}. 
The dynamical temperature $T_d(p)$ is the highest temperature for 
which a ``non-trivial'' solution of the equation exists.
At this temperature ergodicity of the physical dynamics breaks down 
(at least this is what happens in infinite connectivity mean field models)
and we are no longer able to equilibrate the system within an $O(1)$
physical time (i.e. an $O(N)$ computational time).

We looked for a solution of Eqs. (\ref{GeneralRSB_1})-(\ref{GeneralRSB_4})
using the population dynamics algorithm of Ref. \cite{MezardParisiBethe}. 
We checked the ``non-triviality'' of the solution found by considering the
variance of the distributions $\rho(x)$, $\rh(y)$ (more precisely
of the {\it populations} which represent such distributions in the algorithm).

\begin{figure}
\centerline{\hspace{-2cm}
\epsfig{figure=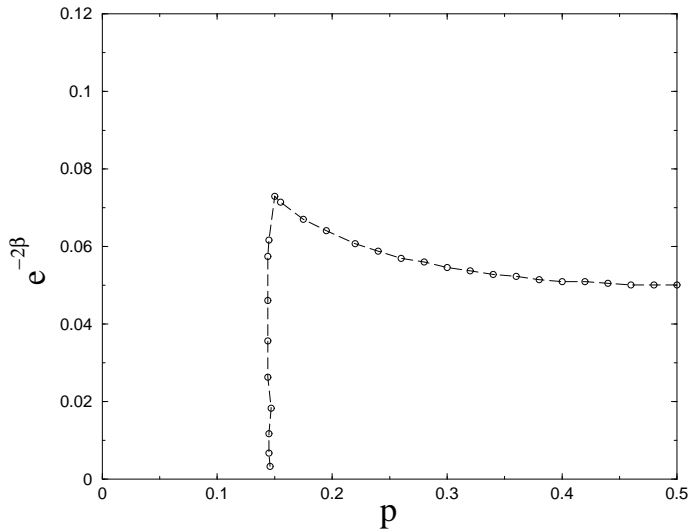,angle=-90,width=0.5\linewidth}}
\caption{The dynamical phase transition for a regular $(6,5)$ code 
(cf. Eq. (\ref{Hamiltonian}) with $k=6$ and $l=5$) with $\zh=1$.}
\label{DynamicalLine}
\end{figure}
We consider the $(6,5)$ regular code because it has well separated 
statical and dynamical thresholds $p_c$ and $p_d$, cf. 
Tab. \ref{ThresholdsTab}.
The resulting dynamical line for the Hamiltonian
(\ref{Hamiltonian}) with $\zh=1$, is reported in Fig. \ref{DynamicalLine}.
The dynamic temperature $T_d(p)$ drops discontinuously 
below a noise $p_d(\zh)$: for $p<p_d(\zh)$ the dynamical 
transition disappears and the system can be equilibrated in linear 
computational time down to zero temperature.
We get $p_d(1) \approx 0.14$, which is in good
agreement with the coding theory results, cf. Tab. \ref{ThresholdsTab}
%
%****************************
%
\subsubsection{Numerical experiments}
\label{NumericalBSC}

We have repeated for the BSC the same kind of simulations already
presented at the end of Sec.~\ref{NumericalBEC} for the BEC.

We have run a set of simulated annealings for the Hamiltonian \ref{Model} of
the (6,5) regular code.  
System size is $N=12000$ and the cooling
rates are the same as for the BEC, the only difference being the
starting and the ending temperatures, which are now $T=1.2$ and
$T=0.2$ (plus a quench from $T=0.2$ to $T=0$ at the end of each
cooling).  This should not have any relevant effect because
$0.2\ll T_d\approx 0.6$.
 
The important difference with respect to the BEC case is that now
we have no fixed spins, all $N$ spins are dynamical variables subject
to a random external field of intensity $h = (1/\beta)\atanh (1-2p)$,
cf. Eq. (\ref{Model}).

\begin{figure}
\begin{tabular}{cc}
\hspace{-1.5cm}
\epsfig{figure=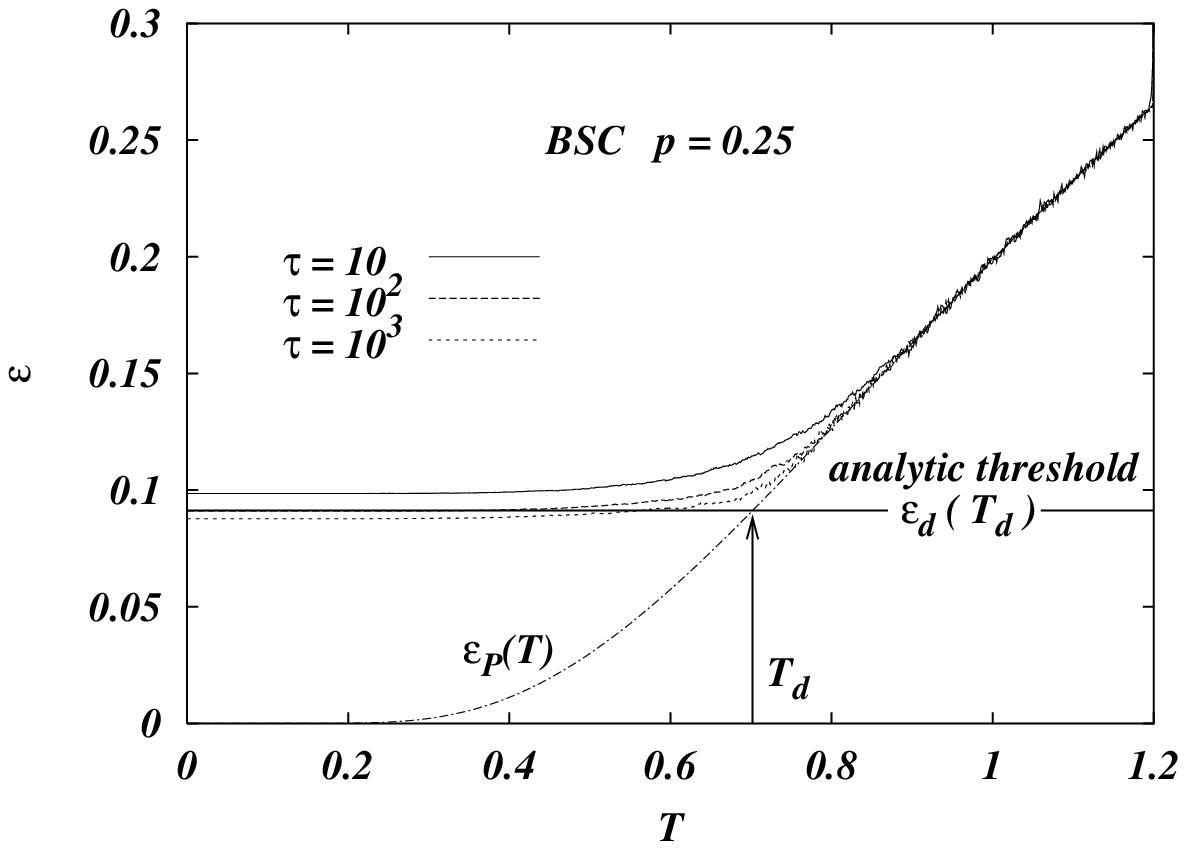,angle=0,width=0.5\linewidth}&
\hspace{0.5cm}
\epsfig{figure=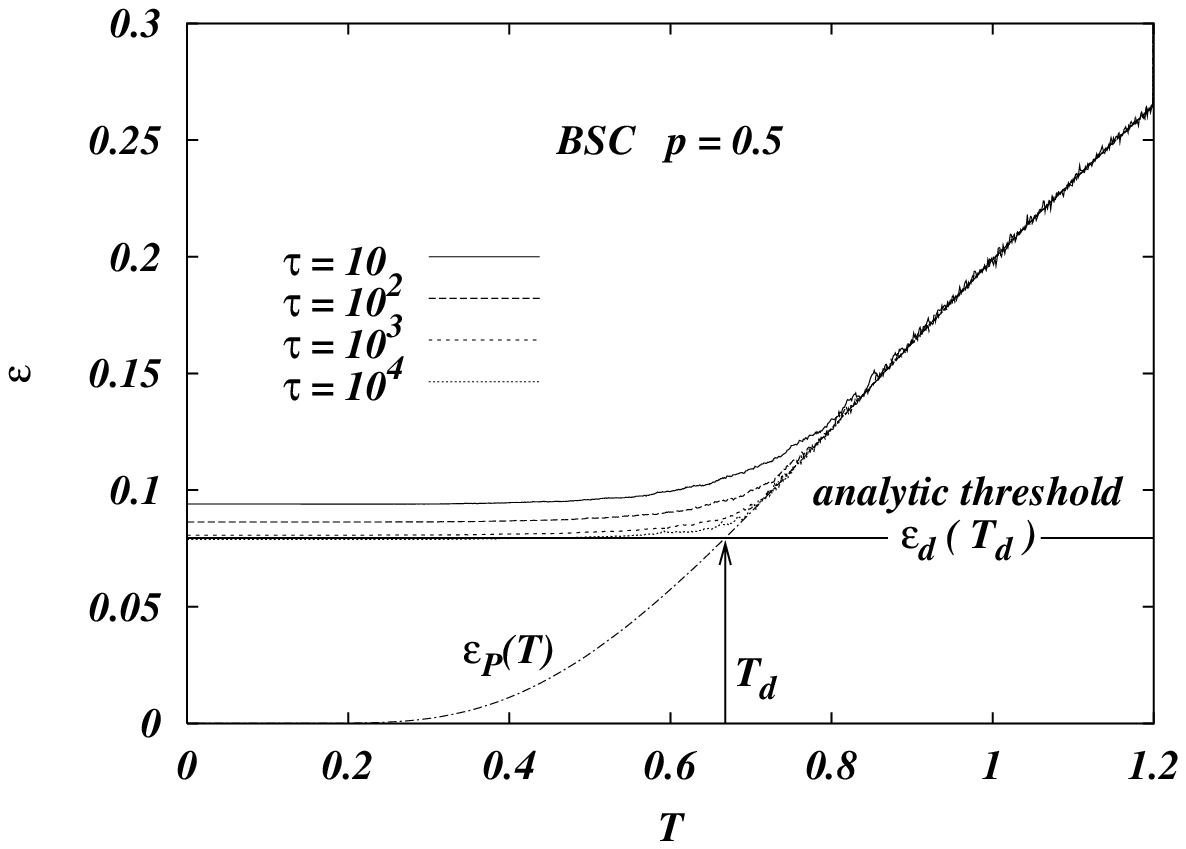,angle=0,width=0.5\linewidth}
\end{tabular}
\caption{Energy relaxation for the Hamiltonian of the (6,5) regular
code during the simulated annealing with $\tau$ MCS per temperature
and 1000 equidistant temperatures in $[0.2,1.2]$. Notice that, in both cases
$p>p_d$. The dot-dashed line is the theoretical prediction for
the paramagnetic exchange energy.}
\label{bsc_ann}
\end{figure}

Also here, as in the case of the BEC, the energy relaxation for
$p>p_d$ undergoes a drastic arrest when the temperature is reduced
below the dynamical transition at $T_d$, see Fig.~\ref{bsc_ann}.

Unfortunately, in this case, we are not able to calculate analytically
the threshold energy $\epsilon_d(0)$, but only the dynamical critical
temperature $T_d$ and then the threshold energy at the transition
$\epsilon_d(T_d)$ which is higher than $\epsilon_d(0)$.  The
difference $\Delta\epsilon = \epsilon_d(T_d) - \epsilon_d(0)$ is
usually not very large (see e.g. the BEC case), but it becomes
apparent when $p$ is decreased towards $p_d$.  Indeed for $p=0.25$
(Fig.~\ref{bsc_ann} left) the Metropolis dynamics is still able to
relax the system for temperatures below $T_d$ and then it reaches an
energy well below $\epsilon_d(T_d)$. On the other hand for $p=0.5$
(Fig.~\ref{bsc_ann} right), where $\Delta\epsilon$ is small the
relaxation below $T_d$ is almost absent and the analytic prediction is
much more accurate. Notice that for this case we have run a still longer 
annealing with $\tau=10^4$: the asymptotic energy is very close to that for
$\tau=10^3$ and hardly distinguishable from the analytical prediction.

\begin{figure}
\centerline{\epsfig{figure=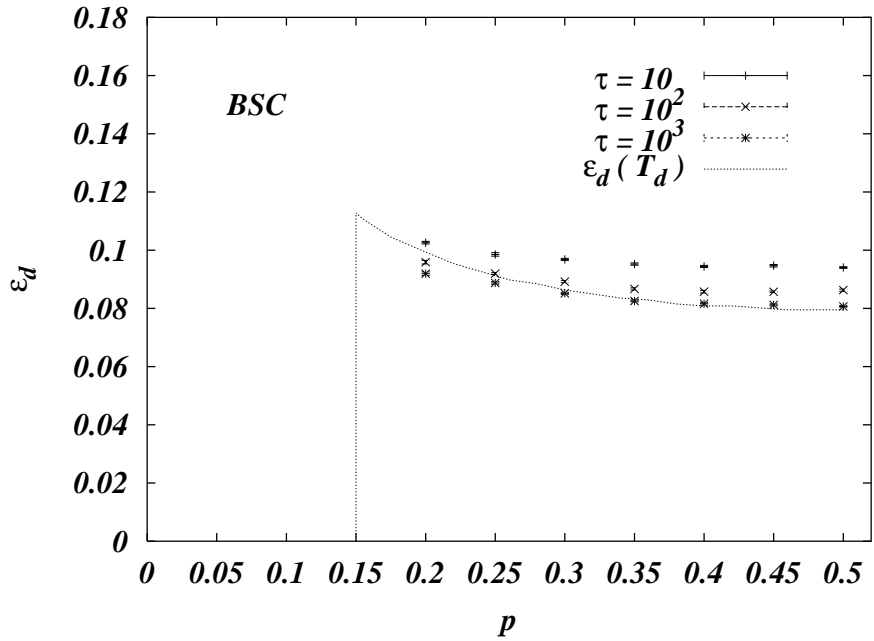,angle=0,width=0.7\linewidth}}
\caption{Lowest energies reached by the simulated annealings. Errors
are sample to sample fluctuations. The theoretical prediction
$\epsilon_d(T_d)$ is computed using the results in Fig. \ref{DynamicalLine}
for $T_d(p)$.}
\label{bsc_e_min}
\end{figure}

In Fig.~\ref{bsc_e_min} we report the lowest energy reached by the
simulated annealing for many values of $p$ and $\tau=10, 10^2, 10^3$,
together with the analytic calculation for the threshold energy at
$T_d$.  This analytical value is an upper bound for the true threshold
energy $\epsilon_d(0)$ where linear algorithms should get stuck, but
it gives very accurate predictions for large $p$ values where
$\Delta\epsilon$ is very small.  In the region of small $p$ a more
complete calculation is needed.
%
%****************************
%
\subsection{Zero temperature}
\label{ZeroTemperatureSection}

This approach follows from a physical intuition that is slightly
different from the one explained in the previous paragraphs.
Once again we will formulate it algorithmically. 
For sake of simplicity we shall refer, in this Section, to the BSC. 
We refer to the Appendix (\ref{GeneralZeroTemperature}) for more general 
formulae.

The overlap between the transmitted codeword and 
the received message 
\begin{eqnarray}
q^{\rm in,out} =\frac{1}{N}\sum_{i=1}^N\sigma^{\rm in}_i\sigma^{\rm out}_i
\, ,
\end{eqnarray}
is, typically, $q^{\rm in, out}=1-2p$. Given the received message, one 
can work in the subspace of all the possible configurations which have 
the prescribed overlap with it\footnote{Of course this is true up to
$O(N^{-1/2})$ corrections. For instance one can work in the space of
configurations $\us$ such that $(1-2p-\delta)N<
\sum_{i=1}^N\sigma_i\sigma_i^{\rm out}<(1-2p+\delta)N$, for some 
small number $\delta$.}, i.e. 
all the  $\us$ such that 
$(1/N)\sum_{i=1}^N\sigma_i\sigma_i^{\rm out}\approx (1-2p)$.
Once this constraint has been imposed (for instance in a Kawasaki-like
Monte Carlo algorithm) one can restrict himself to the exchange part 
of the Hamiltonian (\ref{Hamiltonian}) 
$H_{\rm exch}(\us) = -\sum_k\sum_{(i_1\dots i_k)}
\sigma_{i_1}\cdots\sigma_{i_k}$ and apply the cooling strategy already
described in the previous Section.

Below the static transition $p_c$ there exists a unique codeword 
having overlap $(1-2p)$ with the received signal. This is 
exactly the transmitted one $\us^{\rm in}$. 
This means that $\us^{\rm in}$ is the unique ground state 
of $H_{\rm exch}(\us)$ in the subspace we are considering.
If we are able to keep our system in equilibrium down to $T=0$, the cooling
procedure will finally yield the correct answer to the decoding problem.
Of course, if metastable states are encountered in this process, 
the time required for keeping the system in equilibrium diverges 
exponentially in the size.

We expect the number of such states to be exponentially 
large\footnote{For a related calculation in a fully connected model
see Ref. \cite{Cavagna}.}:
\begin{eqnarray}
{\cal N}_{MS}(\epsilon,q|p)\sim e^{N\Sigma_p(\epsilon,q)}\, ,
\label{ConfEntropy}
\end{eqnarray}
where $\epsilon$ is the exchange energy density $H_{\rm exch}(\us)/N$.
Notice that we emphasized the dependence of these quantities upon the noise
level $p$. In fact the noise level determines the statistics 
of the received message $\us^{\rm out}$.
The static threshold is the noise level at which an exponential number of
codewords with the same overlap as the correct one ($q=1-2p$) appears: 
$\Sigma_p(0,1-2p)>0$.
The dynamic transition occurs where metastable states with the 
same overlap begin to exist: $\Sigma_p(\epsilon,1-2p)>0$ for some 
$\epsilon>0$.
%
%***************************************
%
\subsubsection{The random linear code limit}
\label{ZeroTRLC}

It is quite easy to compute the complexity $\Sigma_p(\epsilon,q)$
in the limit $k,l\to\infty$ with rate $R=1-l/k$ fixed.
In particular, the zeroth order term in a large $k,l$ expansion can be 
derived by elementary methods. 

In this limit we expect the regular $(k,l)$ {\it ensemble} to become identical
to the random linear code (RLC) {\it ensemble}. 
The RLC {\it ensemble} is defined by 
taking each element of the parity check matrix ${\mathbb H}$, cf. Eq. 
(\ref{ParityCheckMatrix}) to be 
$0$ or $1$ with equal probability. Distinct elements are considered to
be statistically independent.

Let us compute the number of configurations $\us$ having a given energy and
overlap with the received message $\us^{\rm out}$.
Given a bit sequence $\utx\neq \ut0$, the probability 
that $L$ out of $M$ equations ${\mathbb H}\utx = \ut0$
are violated is 
\begin{eqnarray}
P_{L,\utx} = \left(\begin{array}{c} M\\L\end{array}\right)2^{-M}\, .
\end{eqnarray}
Therefore the expected number of configurations $\utx$ which violate
$L$ checks and have Hamming distance $W$ from the received message 
$\utx^{\rm out}$ is
\begin{eqnarray}
\overline{{\cal N}_{W,L}} = 
\delta_{W,W_{\utx^{\rm out}}}\delta_{L,0}[1- 2^{-M}]+
\left(\begin{array}{c}N\\ W\end{array}\right)
\left(\begin{array}{c} M\\L\end{array}\right) 2^{-M}\, ,
\end{eqnarray}
where $W_{\utx^{\rm out}}$ is the {\it weight} of $\utx^{\rm out}$, i.e. its
Hamming distance from $\ut0$. Notice that, up to exponentially small 
corrections, the above expression does not depend on $\utx^{\rm out}$.

Introducing the overlap $q=1-2W/N$ and the exchange energy density
$\epsilon = 2L/N$, we get $\overline{{\cal N}_{W,L}}
\sim 2^{N\tilde{\Sigma}(\epsilon,q)}$ with
\begin{eqnarray}
\tilde{\Sigma}(\epsilon,q) = 
{\tt h}[(1-q)/2]+(1-R)\,{\tt h}[\epsilon/2(1-R)]-(1-R)\, .
\label{SigmaRLC}
\end{eqnarray}
The typical number ${\cal N}^{\rm typ}_{W,L}$ of such configurations
can be obtained through 
the usual REM construction: 
${\cal N}^{\rm typ}_{W,L}\sim 2^{N\tilde{\Sigma}(\epsilon,q)}$ when 
$\tilde{\Sigma}(\epsilon,q)\ge 0$ and ${\cal N}^{\rm typ}_{W,L} =0$ otherwise.

Now we are interested in picking, among all the configurations having 
a given energy density $\epsilon$ and overlap $q$, the metastable 
states. In analogy with the REM, this can be done by eliminating 
all the configurations  such that 
$\partial_{\epsilon} \tilde{\Sigma}(\epsilon,q)<0$.
In other words, the number of metastable states is
${\cal N}_{MS}(\epsilon,q)\sim 2^{N\Sigma(\epsilon,q)}$ with
$\Sigma(\epsilon,q)=\tilde{\Sigma}(\epsilon,q)$ when 
$\tilde{\Sigma}(\epsilon,q),\partial_{\epsilon}\tilde{\Sigma}(\epsilon,q)>0$,
$\Sigma(\epsilon,q)=-\infty$ otherwise.

\begin{figure}
\centerline{
\epsfig{figure=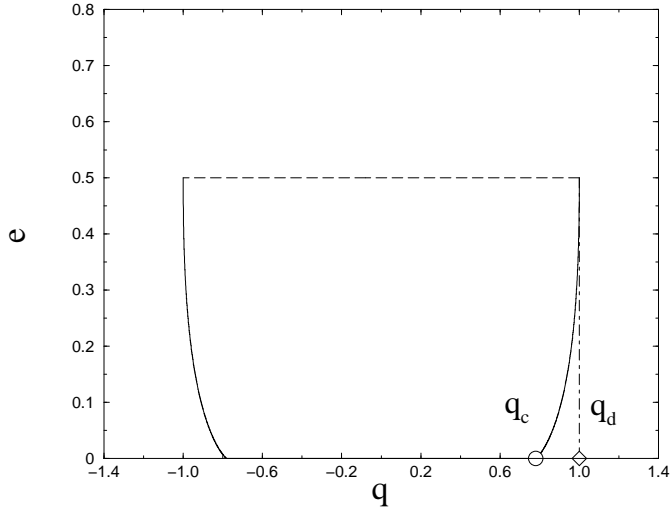,angle=-90,width=0.5\linewidth}}
\caption{Metastable states in the random linear code limit for $R=1/2$:
their number is exponential between the continuous and the dashed lines.
It vanishes discontinuously when the dashed line is crossed and continuously 
when the continuous line is crossed.
The critical and dynamical overlaps are related to the statical and 
critical noise by $q_{c,d}= 1-2p_{c,d}$. In this limit $p_d=0$ and 
$p_c=\delta_{GV}(1/2)\approx 0.110025$.}
\label{ShannonCup}
\end{figure}
In Fig. \ref{ShannonCup} we plot the region of the 
$(\epsilon,q)$ plane for which $\Sigma(\epsilon,q)>0$, for $R=1/2$ codes. 
Notice that, in this limit $\Sigma(\epsilon,q)$ does not depend on the 
received message $\us^{\rm out}$ (and, therefore, is independent of $p$).
As expected we get $p_c=\delta_{GV}(R)$ and $p_d=0$.

\begin{figure}
\begin{tabular}{cc}
\hspace{-0.5cm}
\epsfig{figure=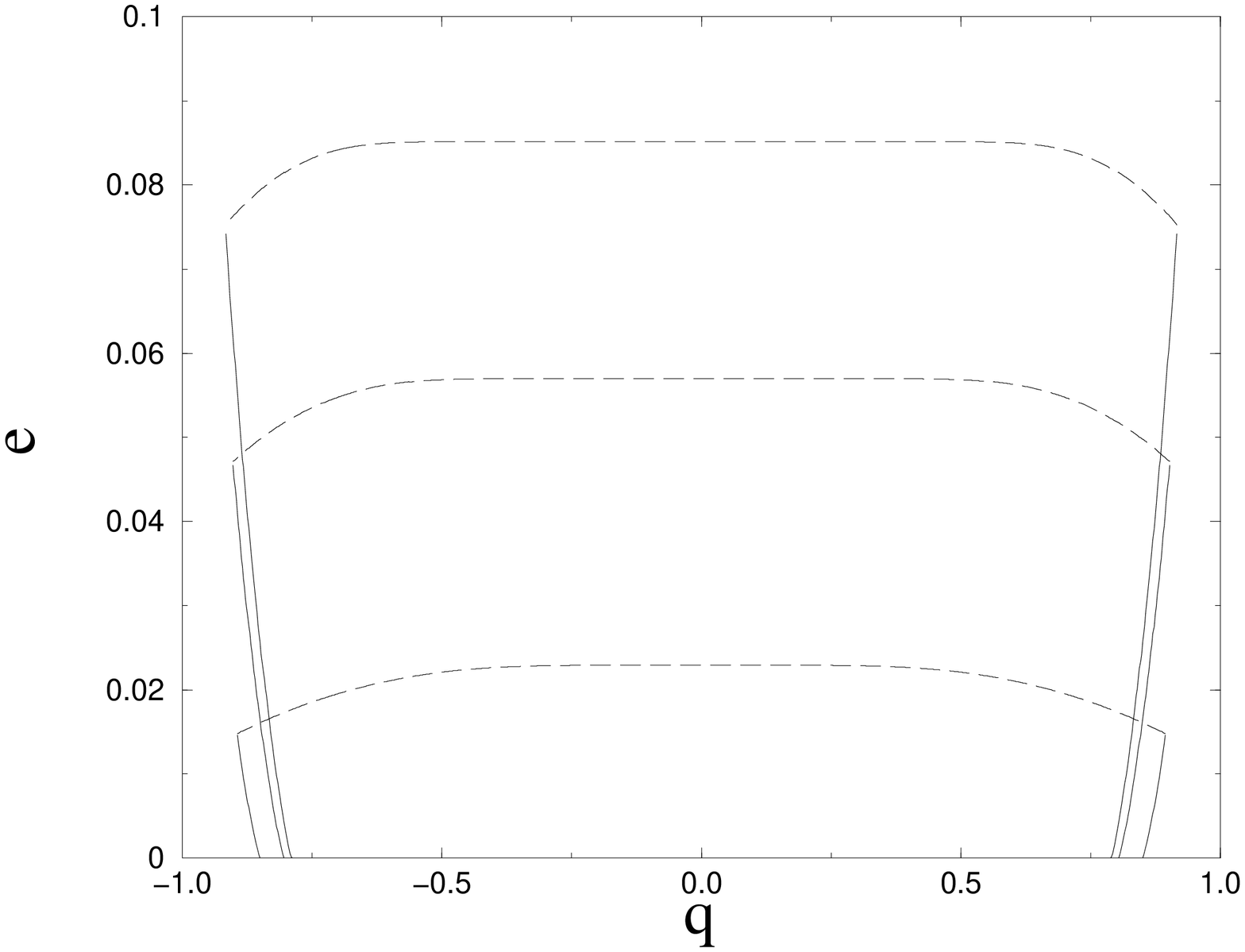,angle=-90,width=0.45\linewidth}&
\epsfig{figure=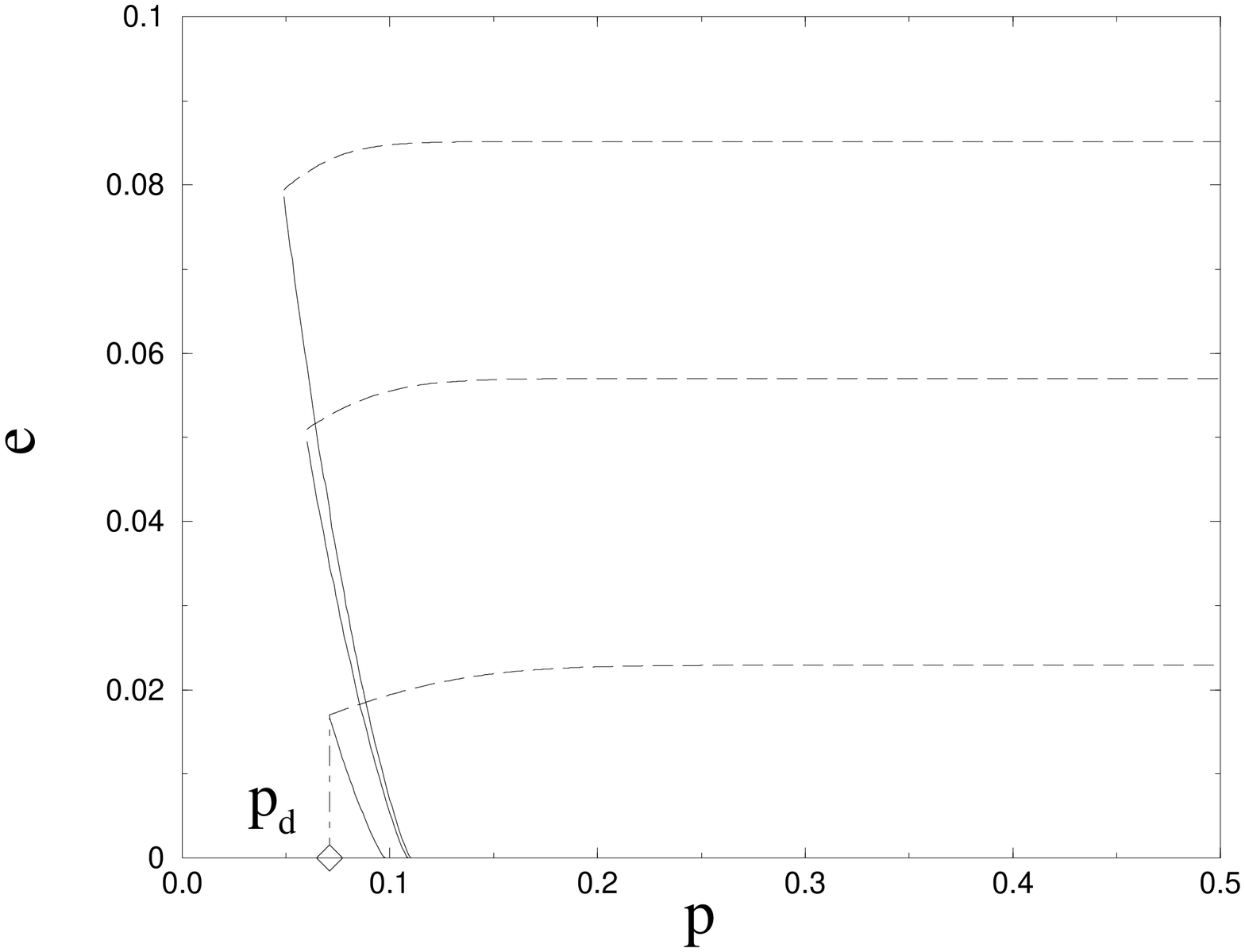,angle=-90,width=0.45\linewidth}
\end{tabular}
\caption{Metastable states for regular $(k,l)$ codes in a large-$k$,$l$
expansion, at fixed rate $R=1/2$. 
We consider (from bottom to top) $(k,l)=(6,3)$, $(10,5)$,
$(14,7)$. On the left we show the region where $\Sigma_{p=0}(\epsilon,q)>0$.
On the right we consider instead $\Sigma_p(\epsilon,1-2p)$.}
\label{Cup_R05}
\end{figure}
In order to get the first non-trivial estimate for the dynamical 
point $p_d$, we must consider the next term in the above expansion.
This correction can be obtained within the replica formalism,
see App. \ref{RLCAppZeroTemp}.
In Fig. \ref{Cup_R05} we reproduce contour of the region 
$\{(\epsilon,q):\Sigma_p(\epsilon,q)>0\}$
for a few regular codes of rate $R=1/2$: $(k,l)=(6,3)$,$(10,5)$,$(14,7)$.
The main difference between these curves and the exact results, cf. 
Sec. \ref{CompleteSection},
is the convexity of the upper boundary of the $\Sigma_p(\epsilon,q)>0$
region (dashed lines in Figs. \ref{ShannonCup} and \ref{Cup_R05}).

The corresponding estimates for $p_c$ and $p_d$ are reported in 
Tab. \ref{ExpansionTable}.
\begin{table}
\centerline{
\begin{tabular}{|c|c|c|}
\hline
$(k,l)$ & $p_c$ & $p_d(1)$ \\
\hline
\hline
$(6,3)$  & $0.097$ & $0.071$ \\
\hline
$(10,5)$ & $0.108$ & $0.060$\\
\hline
$(14,7)$ & $0.109$ & $0.049$\\
\hline
$(6,5)$  & $0.264$ & $0.108$\\
\hline
\end{tabular}}
\caption{Dynamical and statical thresholds at the first nontrivial 
order in a large $k$,$l$ expansion, cf. Tab. \ref{ThresholdsTab}.}
\label{ExpansionTable}
\end{table}
%
%
%***************************************
%
\subsubsection{The complete calculation}
\label{CompleteSection}

\begin{figure}
\centerline{\epsfig{figure=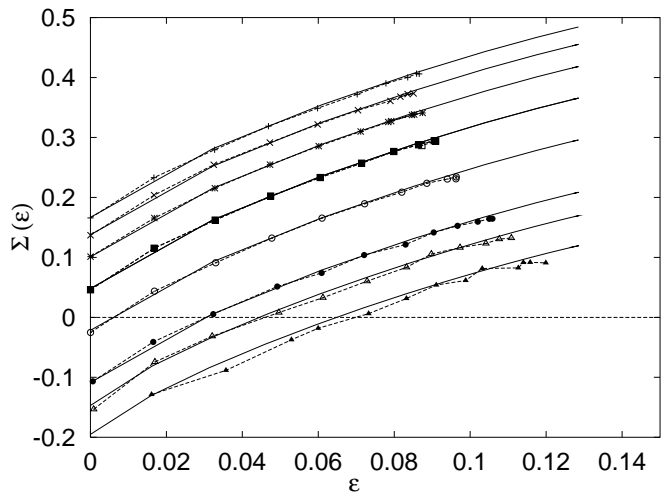,angle=-90,width=0.5\linewidth}}
\caption{The configurational entropy versus the energy for 
the $(6,5)$ regular code. Symbols refer to various noise levels.
From top to bottom $p=0.5, 0.4, 0.35, 0.3, 0.25, 0.2, 0.18, 0.155$.
Continuous lines give the result of a variational computation,
cf. App. \ref{VariationalAppendix}.}
\label{figure2}
\end{figure}

The full 1RSB solution for can be obtained through the
population dynamics method \cite{MezardParisiBethe}. 
Here, as in Sec. \ref{CompleteFiniteBeta}, 
we focus on the example of the $(6,5)$ code.  
In Fig. \ref{figure2} we plot the configurational entropy as a function
of the energy of the states along the lines of constant $q$, together with the
corresponding results obtained within a simple variational approach,
cf. App. \ref{VariationalAppendix}. 
The approximate treatment is in quantitative agreement with the 
complete calculation for $\epsilon<\epsilon_d$, but predicts a 
value for the threshold energy which is larger
than the correct one: $\epsilon_d^{var} > \epsilon_d$. 
Here $\epsilon_d^{var} \approx 0.127$ and almost $p$-independent.

Unhappily the estimate of the dynamic energy obtained 
from this curves is not very precise. Moreover, at least two more 
considerations prevent us from comparing these results with the ones of
simulated annealing simulations, cf. Sec. \ref{NumericalBSC}: 
$(i)$ In our annealing experiments the overlap with the received message 
$\us^{\rm out}$ is  free to fluctuate; 
$(ii)$ We cannot exclude the 1RSB solution 
to become unstable at low temperature.

However the population dynamics solution give the estimate 
$p_d\lesssim 0.155$.
This allows us to confirm that the 
point $p_d=0.139$ where the decoding algorithm fails to decode, 
cf. Tab. \ref{ThresholdsTab}, coincides with the point where the metastable
states appear.
%
%*********************************************************************
%
\section{Conclusion}
\label{ConclusionSection}

We studied the dynamical phase transition for a large class of diluted
spin models in a random field, the main motivation being their correspondence
with very powerful error correcting codes.

In a particular case, we were able
to show that the dynamic critical point coincides exactly with
the critical noise level for an important class of decoding algorithms,
cf. Sec. \ref{BECSection} and App. \ref{BECAppendix}.
For a general model of the noisy channel, we couldn't present
a completely explicit proof of the same statement. 
However, within numerical precision, we obtain identical values for
the algorithmic and the statistical mechanics thresholds.

It may be worth listing a few interesting problems which emerge from our 
work:
\begin{itemize}
\item Show explicitly that the identity between statistical mechanics and 
algorithmic thresholds holds in general. From a technical point of view, this 
is a surprising fact because the two thresholds are obtained, respectively,
within a replica symmetric, cf Eqs. (\ref{DensityEvolution_1}),
(\ref{DensityEvolution_2}), and a one-step replica symmetry 
breaking calculations.
\item We considered message-passing and simulated annealing algorithms.
Extend the above analysis to other classes of algorithm (and, 
eventually, to any linear time algorithm).
\item Message passing decoding algorithms get stuck because they are unable
to decode some fraction of the received message, the ``hard'' bits, while
they have been able to decode the other ones, the ``easy'' bits, cf. 
App. \ref{BeyondSection}.
A closer look at this heterogeneous behavior would be very fruitful.
\end{itemize}
\vspace{0.5cm}

It is a pleasure to thank R.~Zecchina who participated to the early 
stages of this work.
\appendix
%
%*********************************************************************
%
\section{Calculations: binary erasure channel}
\label{BECAppendix}

In this Appendix we give the details of the replica calculation for
the BEC. Notice that although we use the regular $(6,3)$
code as a generic example, all the computations are presented for
general degree distributions $\{c_k\}$ and $\{v_l\}$.
%
%*******************************************************************
%
\subsection{Replica symmetric approximation}
The replica symmetric calculation is correct \cite{GallagerAM} as long as we
focus on codewords (i.e. on zero energy configurations). 
The main reason is the Nishimori symmetry 
\cite{Nishimori1,Nishimori2,Nishimori3} which holds at $\beta=\infty$
and $\zh=1$ for the model (\ref{Model}).
Therefore the replica symmetric approximation gives access
the correct noise level $p_c$ for the statical
phase transition. Although such computations have been already 
considered in Refs. \cite{SaadRegular,GallagerAM}, 
it is interesting to review them for the BEC, which
allows for a cleaner physical interpretation. Moreover here 
we generalize the already published results by considering a generic irregular 
construction.

We parametrize the order parameters $\l(\vs)$, $\lh(\vs)$, cf. Eq.
(\ref{Action}) using the Ansatz
\begin{eqnarray}
\l(\vs) = \int\! d\pi(x) \frac{e^{\beta x\sum_a\sigma^a}}
{(2\cosh \beta x)^n}\, ,\;\;\;\;\;\;\;\;
\lh(\vs) = \int\! d\ph(y) \frac{e^{\beta y\sum_a\sigma^a}}
{(2\cosh \beta y)^n}\, ,
\end{eqnarray}
where we adopted the notation $\int\! d\pi(x)\, (\cdots)\equiv
\int\! dx\,\pi(x)\, (\cdots)$. It is easy to see 
that the order parameters have the form
\begin{eqnarray}
\pi(x) =  (1-p)\delta_{\infty}(x) +p\, \rho(x)\, ,\;\;\;\;\;\;\;
\ph(y) =\rh(y)\, ,
\end{eqnarray}
where $\delta_{\infty}(\cdot)$ is a Dirac delta function at $+\infty$,
and $\rho(x)$, $\rh(y)$ are supported on finite effective fields.
Physically we are distinguishing the sites which correspond to 
correctly received bits (and on which an infinite magnetic field acts)
from the other ones.

The new order parameters $\rho(x)$ and $\rh(y)$ satisfy
\begin{eqnarray}
\rho(x) & = & \frac{1}{\la}\sum_{l=2}^{\infty} v_l l\,
\int\!\prod_{i=1}^{l-1}d\rh(y_i)\, \delta(x-y_1-\dots-y_{l-1})\, ,
\label{Rho}\\
\rh(y) & = & \sum_{\nu = 0}^{\infty} f_{\nu}
\int\!\!\prod_{i=1}^{\nu}d\rho(x_i)\,\delta(y-
\frac{1}{\beta}\atanh[\tanh\beta \tanh\beta x_1\cdots\tanh\beta
x_{\nu}])\, ,\label{RhoHat}
\end{eqnarray}
where 
\begin{eqnarray}
f_{\nu} = \frac{1}{\ka}\sum_{k=\nu+1}^{\infty}c_k k\left(
\begin{array}{c} k-1\\ \nu\end{array}\right)p^{\nu}(1-p)^{k-1-\nu}\, .
\label{fnudefinition}
\end{eqnarray}
It is useful to introduce the generating function $f(x)$ of
the coefficients $\{f_{\nu}\}$:
$f(x)\equiv\sum_{\nu=0}^{\infty}f_{\nu}x^{\nu}$.
It is easy to show that $f(x) = c'(1-p+px)/c'(1)$.

The replica symmetric free energy is obtained by substituting the above Ansatz
in Eq. (\ref{Model}):
\begin{eqnarray}
\beta \phi[\rho,\rh] & =& \la p \int\!d\rho(x)\!\int\!d\rh(y) \,
\log[1+\tanh \beta x\tanh \beta y]-\\
&&-\frac{\la}{\ka}\sum_{\nu=0}^{\infty}g_{\nu}
\int\!\prod_{i=1}^{\nu}d\rho(x_i)\, \log[1+\tanh\beta\tanh\beta x_1
\dots \tanh\beta x_{\nu}]-\nonumber\\
&&\!\!\!\!\!\!\!\!\!\!\!\!\!\!\!\!\!\!
-p \sum_{l=2}^{\infty}v_l\int\!\prod_{i=1}^{\nu}d\rh(y_i)\, 
\log[\prod_{i=1}^{l}(1+\tanh\beta y_i)+\prod_{i=1}^{l}(1-\tanh\beta
y_i)]
-\frac{\la}{\ka}\log\left(\frac{1+e^{-2\beta}}{2}\right)\, ,
\nonumber
\end{eqnarray}
with
\begin{eqnarray}
g_{\nu} \equiv \sum_{k=\nu}^{\infty} c_k \left(
\begin{array}{c} k\\ \nu\end{array}\right)p^{\nu}(1-p)^{k-\nu}\, .
\end{eqnarray}
The generating function of the coefficients $\{g_{\nu}\}$
is given by $g(x) = c(1-p+px)$.
Notice that $\{ g_{\nu}\}$ is the effective degree distribution of 
parity check nodes (i.e. the analogous of $\{c_k\}$), 
once the received bits have been eliminated.

Let us now consider the $\beta\to\infty$ limit. We look for solution 
of the following form (to the leading order):
\begin{eqnarray}
\rho(x) = \sum_{q=-\infty}^{+\infty}\rho_q\, \delta(x-q)\, ,
\;\;\;\;\;\;\;\;
\rh(y) = \rh_+\delta(y-1)+\rh_0\delta(y)+\rh_- \delta(y+1)\, .
\label{BetaInftyAnsatz}
\end{eqnarray}
If we define $\rho_+ \equiv \sum_{q>0}\rho_q$ and
$\rho_- = \sum_{q<0}\rho_q$, it is easy to get 
four coupled equations for the four variables $\rho_{\pm}$ and 
$\rh_{\pm}$:
\begin{eqnarray}
\rh_+ & = & \frac{1}{2}[f(\rho_++\rho_-)+f(\rho_+-\rho_-)]\, ,
\label{BetaInfty1}\\
\rh_- & = & \frac{1}{2}[f(\rho_++\rho_-)-f(\rho_+-\rho_-)]\, ,
\label{BetaInfty2}\\
\rho_+ & = & \frac{1}{\la}\sum_{l=2}^{\infty} v_l l\sum_{n_+>n_-,n_0} 
\frac{(l-1)!}{n_+!n_0!n_-!}\, \rh_+^{n_+}\rh_0^{n_0}\rh_-^{n_-}\
\, \delta_{n_++n_0+n_-,l-1}\, ,\\
\rho_- & = & \frac{1}{\la}\sum_{l=2}^{\infty} v_l l\sum_{n_->n_+,n_0} 
\frac{(l-1)!}{n_+!n_0!n_-!}\, \rh_+^{n_+}\rh_0^{n_0}\rh_-^{n_-}
\, \delta_{n_++n_0+n_-,l-1}\, .\label{BetaInfty4}
\end{eqnarray}
In these equations $\rh_0$ should be regarded as a shorthand for
$1-\rh_+-\rh_-$. The complete distribution of the fields can be 
reconstructed from $\rh_{\pm}$ using the equation below
\begin{eqnarray}
\rho_q = \frac{1}{\la} \sum_{l=2}^{\infty} v_l l\sum_{n_+,n_0,,n_-} 
\frac{(l-1)!}{n_+!n_0!n_-!}\, \rh_+^{n_+}\rh_0^{n_0}\rh_-^{n_-}
\, \delta_{n_++n_0+n_-,l-1}\delta_{n_+-n_-,q}\, .
\end{eqnarray}

\subsubsection{Ferromagnetic solutions}

It is clear that Eqs. (\ref{BetaInfty1})-(\ref{BetaInfty4}) admit
solutions with $\rho_- = \rh_- = 0$. 
Defining $\rho \equiv p(1-\rho_+)$ and $\rh \equiv 1-\rh_+$, 
and using Eqs. (\ref{BetaInfty1})-(\ref{BetaInfty4}), we get
\begin{eqnarray}
\rh   =  1-\frac{c'(1-\rho)}{c'(1)}\, ,\;\;\;\;\;\;\;\;
\rho  =  p\frac{v'(\rh)}{v'(1)}\, .
\label{SaddleFerro}
\end{eqnarray}
The energy of such a solution is always zero.
This means that there exists always at least one codeword which is 
compatible with the received message (this is true by construction).

In order to compute the entropy (and therefore the number of 
such codewords), the finite-temperature corrections 
to the Ansatz (\ref{BetaInftyAnsatz}) must be computed.
More precisely we write:
\begin{eqnarray}
\rho(x) = \sum_{q=0}^{+\infty} \rho_q \,
\beta\,u_q(\beta(x-q))\, ,\;\;\;\;\;
\rh(y) = \rh_0\,\beta\,\uh_0(\beta y) + \rh_1\,\beta\, \uh_1(\beta
(y-1))\, ,
\end{eqnarray}
where $u_q(\cdot)$ and $\uh_q(\cdot)$ are normalized distributions
centered in zero with a well-behaved $\beta\to\infty$ limit.
Using this Ansatz in Eqs. (\ref{Rho})-(\ref{RhoHat}) and taking the
$\beta\to\infty$ limit one obtain two coupled 
equations for $u_0(\cdot)$ and $\uh_0(\cdot)$. 
These equations can be studied using a population
dynamics algorithm. The outcome is $u_q(x)=\uh_q(x) = \delta(x)$. 
The entropy is therefore correctly given by the simple Ansatz 
(\ref{BetaInftyAnsatz}). The result is reported in 
Eq. (\ref{ParaEntropy}).

\subsubsection{Glassy solutions}

Now we look for solutions of the saddle point equations
of the form (\ref{BetaInftyAnsatz}) with 
$\rho_-,\, \rh_- >0$. Such solutions have positive energy and will
correspond (at most) to metastable states.
The energy is easily written in terms of $\rho_+, \rho_-, \rh_+,
\rh_-$ ($\rh_0$ has to be interpreted as a shorthand for 
$1-\rh_+-\rh_-$):
\begin{eqnarray}
\epsilon[\rho,\rh] &= &-2p\la(\rho_+\rh_-+\rho_-\rh_+)+
\frac{\la}{\ka}\left\{c[1-p+p(\rho_++\rho_-)]-c[1-p+p(\rho_+-\rho_-)]\right\}+
\nonumber\\
&&+2p\sum_{l=2}^{\infty}v_l\sum_{n_+,n_0,n_-}\!\!\!\!\!\!
{}'\,\frac{l!}{n_+!n_0!n_-!}
\rh_+^{n_+}\rh_0^{n_0}\rh_-^{n_-}\min(n_+,n_-)\, ,
\label{RSEnergy}
\end{eqnarray}
where the sum $\sum'$ is intended to be carried over the integers 
$n_+,n_0,n_- \ge 0$ such that $n_++n_0+n_-=l$.

Notice that $\rh_+$ and $\rh_-$ can be unambiguously eliminated from
the above expression by making use of Eqs. (\ref{BetaInfty1}),
(\ref{BetaInfty2}). We are then left with a function of two variables:
$\epsilon(\rho_+,\rho_-)$.
Rather than studying such a function for general degree distributions 
$\{ c_k \}$ and $\{ v_l \}$, we shall focus on the regular 
$(6,3)$ case: this corresponds to using $c(x) = x^6$ and $v(x) = x^3$
in Eq. (\ref{RSEnergy}). 
We expect that the behavior found in this case is generic.
\begin{figure}
\begin{tabular}{ccc}
\epsfig{figure=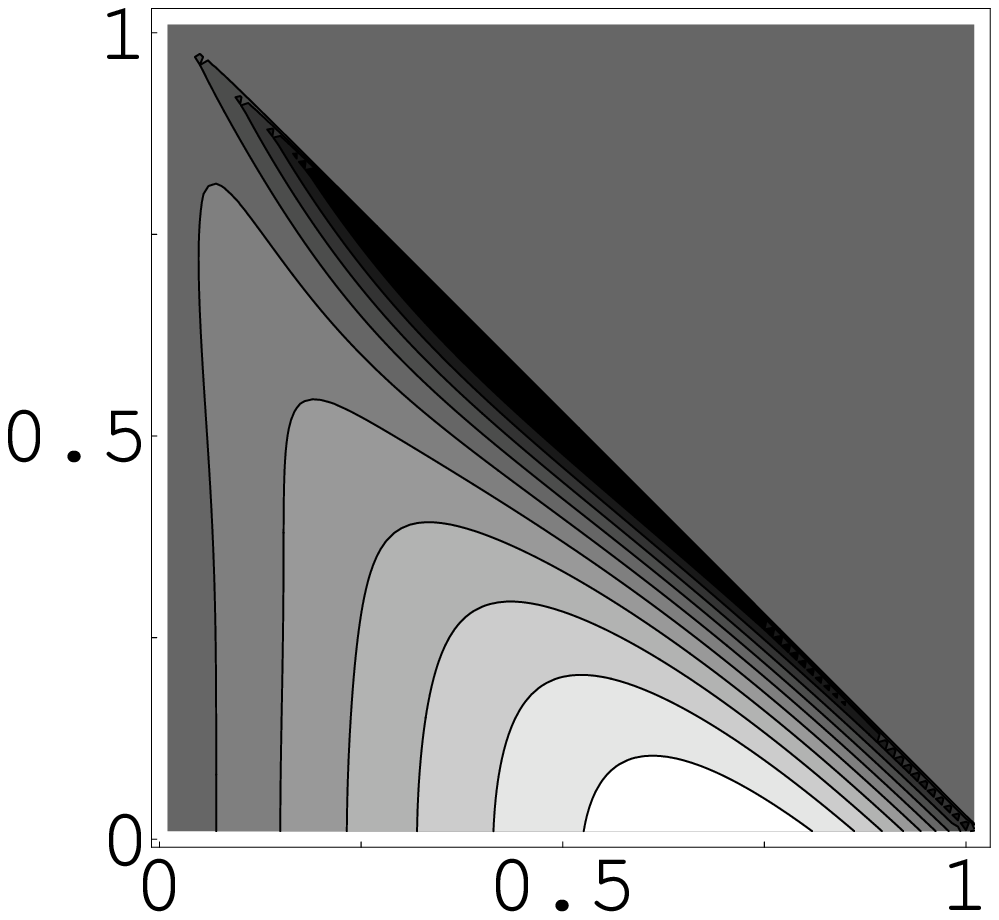,width=0.3\linewidth}
&\epsfig{figure=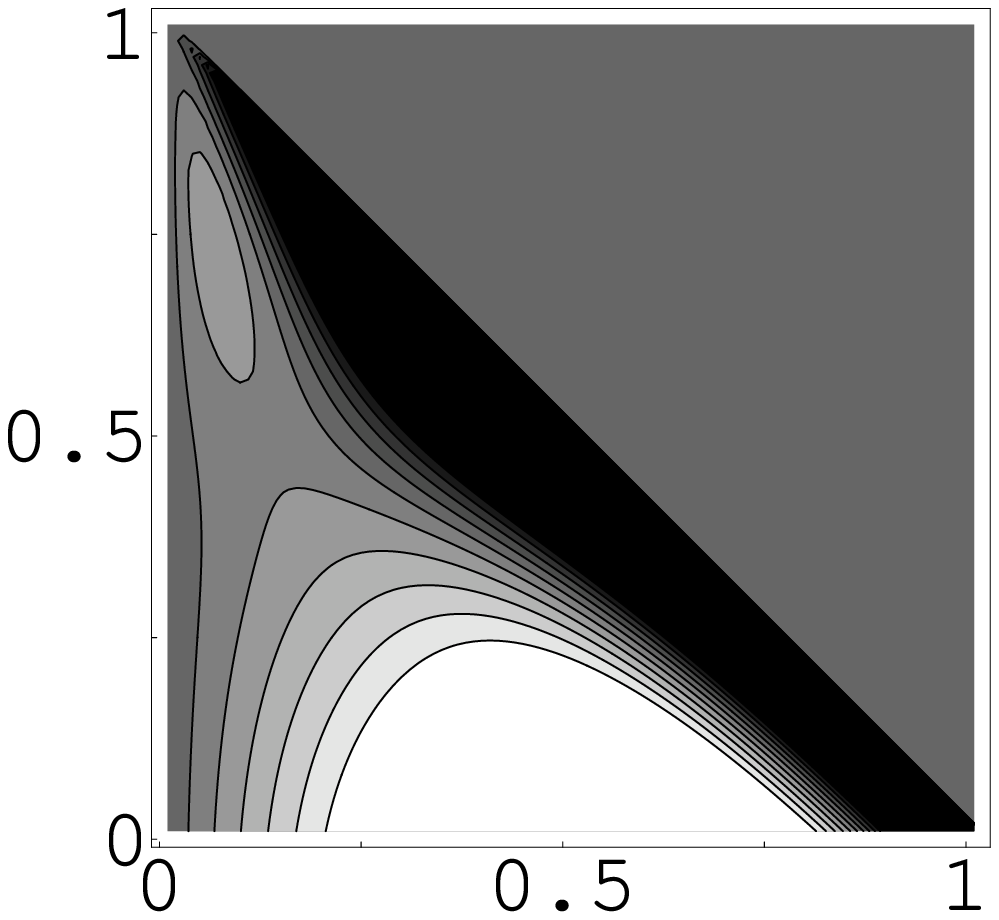,width=0.3\linewidth}
&\epsfig{figure=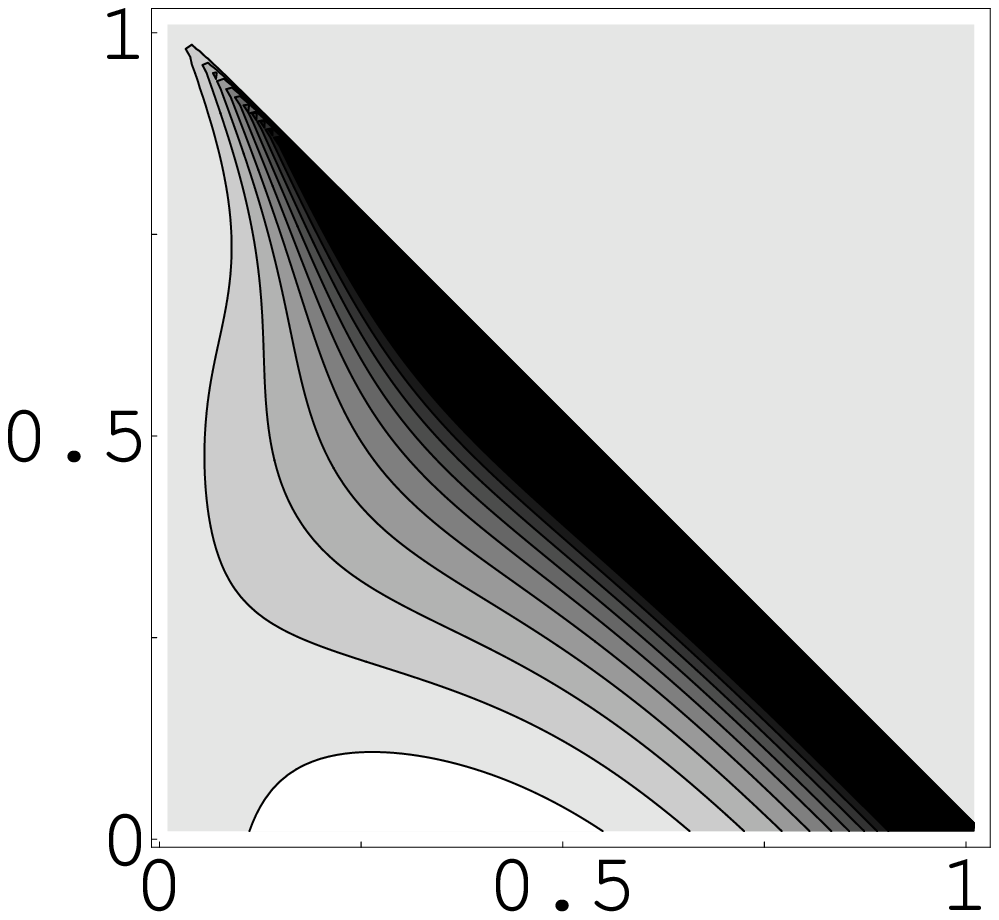,width=0.3\linewidth}
\end{tabular}
\caption{The replica symmetric energy (\ref{RSEnergy}) as a function
of $\rho_+$ (vertical axis) and $\rho_-$ (horizontal axis),
for $p=0.35$, $0.4$ and $0.45$ (from left to right).
Notice that only the region $\rho_++\rho_-\le 1$ is meaningful.
The energy vanishes as $\rho_-\to 0$.}
\label{Contour}
\end{figure}

In Fig. \ref{Contour} we plot $\epsilon(\rho_+,\rho_-)$ for three
different values of the erasure probability $p_1$, $p_2$ and $p_3$, 
with $p_1<p_2<p_d<p_3$. It is easy to guess the qualitative behavior
of $\epsilon(\rho_+,\rho_-)$ as $p$ is varied. For small values of $p$
no glassy extremal point with $\rho_+,\rho_->0$ can be found. At some
value $p^*<p_d$ two such points appear: a maximum and a saddle.
At the dynamical threshold $p_d$ the saddle point collapses onto the 
$\rho_-=0$ axis.

\begin{figure}
\begin{tabular}{cc}
\epsfig{figure=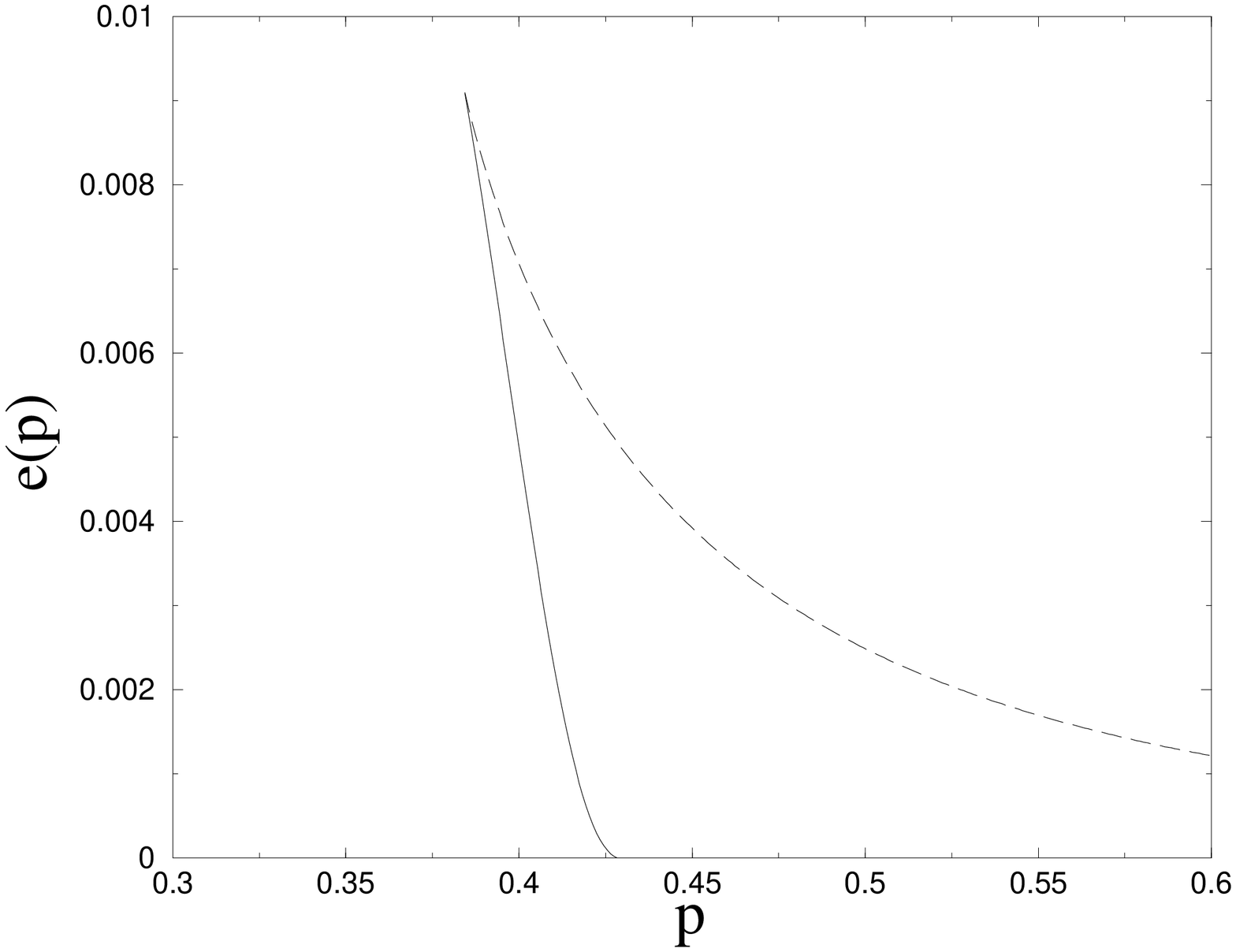,angle=-90,width=0.45\linewidth}
&\epsfig{figure=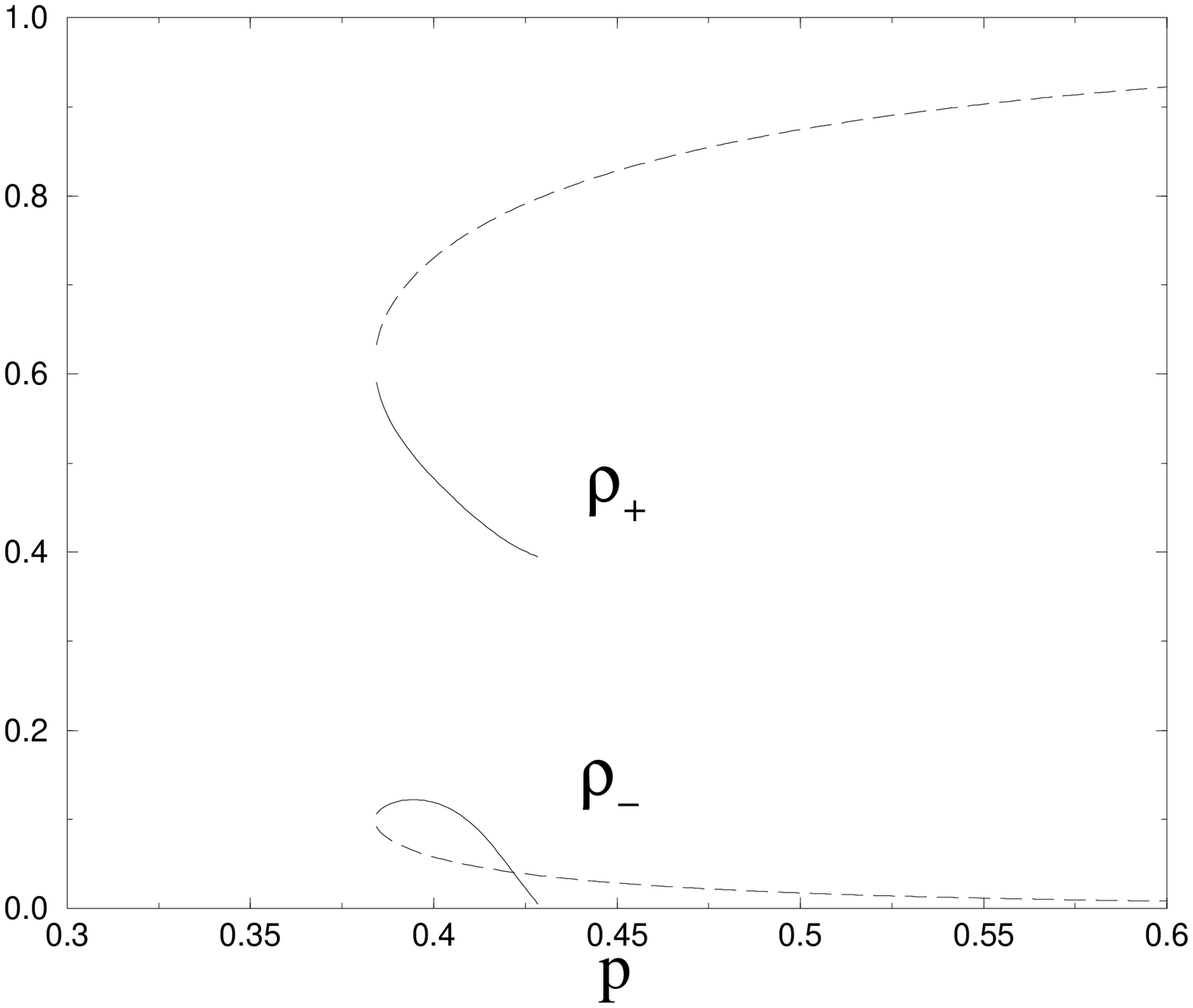,angle=-90,width=0.4\linewidth}
\end{tabular}
\caption{The energy density (right-hand graph) and the position in the
$(\rho_+,\rho_-)$ plane of the two non-trivial RS solutions as a
function of the erasure probability $p$. The continuous lines refer 
to the saddle point and the dashed line to the maximum.}
\label{EdipRS}
\end{figure}
This picture can be confirmed by more careful study. The two glassy
solutions appear at $p^* \approx 0.3844$. In Fig. (\ref{EdipRS}) we report the
corresponding energies and their position in the $(\rho_+,\rho_-)$
plane as functions of $p$.

Which of the two solutions is the physical one? We argue in favor of
the saddle point for the following reasons:
\begin{itemize}
\item According to the standard recipe \cite{SpinGlass}, free energy has to
be minimized with respect to overlaps involving an odd number of
replicas and maximized with respect to overlaps of an even number of replicas. 
In our case $\rho_+-\rho_-$ is the magnetization (an odd overlap) and
$\rho_++\rho_-$ is related to the two replicas overlap. We expect
therefore the physical solution to have a stable and an unstable directions.
\item Two well-known relatives of our model are the ferromagnet and
the $\pm J$ spin glass on random hypergraphs 
\cite{FranzEtAlExact,FranzEtAlFerromagn}. In this cases a
symmetric solution ($\rho_+=\rho_-$) exists and is accepted to
describe the glassy states.
When regarded in the full $(\rho_+,\rho_-)$ plane this solution appear
to be a saddle point in the spin glass model, and to have zero 
second derivative in the ferromagnetic model.
\item One can try to solve Eqs, (\ref{BetaInfty1})-(\ref{BetaInfty4})
iteratively. This procedure is analogous, within the Ansatz
(\ref{BetaInftyAnsatz}), to
the population dynamics algorithm,
which is commonly believed to converge to the correct solution. The iterative
procedure converges to the saddle point. 
\end{itemize}

Nonetheless the qualitative picture that one would expect from the
behavior of decoding algorithms is quite different from the one
offered by the replica symmetric solution. 
We expect metastable state to appear with positive energy at $p_d$ and
the minimum among their energies to vanish at $p_c$.
It is therefore necessary to go beyond the replica symmetric approximation.
%
%*******************************************************************
%
\subsection{Replica symmetry breaking}
\label{BEC_RSB}
The exact computation of the 1RSB free energy is a very difficult task
for a finite connectivity model \cite{MonassonRSB}.
Good results can be obtained the following variational Ansatz
(see Ref. \cite{BiroliVariational} for the general philosophy of the 
variational approach)
\begin{eqnarray}
\l(\vs) & = & (1-p)\delta_{\vs,\vs_0}+p f(\us^{(1)})\cdot\dots\cdot
f(\us^{(n/m)})\, ,\\
\lh(\vs) & = & \fh(\us^{(1)})\cdot\dots\cdot\fh(\us^{(n/m)}) 
\end{eqnarray}
where $\us^{(\alpha)} = 
(\sigma^{(\alpha-1)m+1},\dots,\sigma^{\alpha m})$. 
This amounts to considering a fraction the spins (namely, the ones with an
infinite magnetic field) as frozen in the $+1$ state, and assuming
all the other spins to be equivalent. 
In the $n\to 0$ limit we get $\partial_n S[\l,\lh]\to \phi[f,\fh]$ with
\begin{eqnarray}
\phi[f,\fh] &= & \frac{\la
p}{m}\log\left(\sum_{\us}f(\us)\fh(\us)\right)-
\frac{p}{m}\sum_{l=2}^{\infty}v_l\log\left(\sum_{\us}\fh(\us)^l\right)-\\
&&-\frac{\la}{\ka m}\sum_{\nu = 0}^{\infty}g_{\nu} 
\log\left[\sum_{\us_1\dots\us_k}
J^{(m)}_{\beta}(\us_1,\dots,\us_k)f(\us_1)\dots
f(\us_k)\right]\nonumber\, ,
\label{FactorizedFreeEnergy}
\end{eqnarray}
where $\us$ are $m$-components replicated spins and
Notice that the energy (\ref{FactorizedFreeEnergy}) is invariant under
a multiplicative rescaling of $f(\us)$ and $\fh(\us)$. We shall fix
this freedom by requiring that $\sum_{\us} f(\us)=\sum_{\us}\fh(\us) = 1$.

Substituting
\begin{eqnarray}
f(\us) \equiv \int \! dx \, \rho(x) \,
\frac{e^{\beta x\sum_{a=1}^m\sigma^a}}{(2 \cosh\beta x)^m}\, , \;\;\;\;\;\;\;\;
\fh(\us) \equiv \int \! dy \, \rh(y) 
\frac{e^{\beta y\sum_{a=1}^m\sigma^a}}{(2 \cosh\beta y)^m}\, ,
\end{eqnarray}
we obtain
\begin{eqnarray}
\beta\phi[\rho,\rh] & = & \frac{\la p}{m} \log\left[
\int\!d\rho(x)d\rh(y)\, (1+\tanh\beta x\tanh\beta y)^m\right]-
\frac{\la}{\ka}\log\left(\frac{1+e^{-2\beta}}{2}\right)-\nonumber\\
&&\!\!\!\! -\frac{\la}{\ka m}\sum_{\nu=0}^{\infty}g_{\nu}
\log\left[\int\!\prod_{i=1}^{\nu}d\rho(x_i)\,(1+\tanh\beta 
\tanh\beta x_1\cdots\tanh\beta x_{\nu})^m\right]-\\
&&\!\!\!\! -\frac{p}{m}\sum_{l=2}^{\infty}v_l
\log\left\{\int\!\prod_{i=1}^{l}d\rh(y_i)\,
\left[\prod_{i=1}^l(1+\tanh\beta y_i)+
\prod_{i=1}^l(1-\tanh\beta y_i)\right]\right\}\, ,\nonumber
\end{eqnarray}
and the corresponding saddle point equations:
\begin{eqnarray}
\frac{\rho(x)}{(2\cosh\beta x)^m} & = & \frac{1}{{\mathbb Z}\,\la}
\sum_{l=2}^{\infty}
v_l l\, B_l^{-1}\,\int\!\prod_{i=1}^{l-1}\frac{d\rh(y_i)}{(2\cosh\beta y_i)^m}\,
\delta(x-\sum_{i=1}^{l-1}y_i)\, ,\\
\rh(y) & = & \frac{1}{{\mathbb Q}}\sum_{\nu=1}^{\infty}f_{\nu-1}A^{-1}_{\nu}
\int\!\prod_{i=1}^{\nu-1}d\rho(y_i)\,\delta(y-\frac{1}{\beta}
\atanh[\tanh\beta\tanh\beta y_1\cdots\tanh\beta y_{\nu-1})\, ,\nonumber\\
\end{eqnarray}
where $f_{\nu-1}\equiv g_{\nu}\nu/(p \ka)$, cf. Eq. (\ref{fnudefinition}), and 
\begin{eqnarray}
B_l & \equiv & \int\!\prod_{i=1}^{l}d\rh(y_i)\, 
\left[\prod_{i=1}^l(1+\tanh\beta y_i)+\prod_{i=1}^l(1-\tanh\beta y_i)
\right]^m\\
A_{\nu}& \equiv& \int\!\prod_{i=1}^{\nu}d\rho(x_i)\,
[1+\tanh\beta\tanh\beta x_1\cdots\tanh\beta x_{\nu}]^m\, .
\end{eqnarray}
The constants ${\mathbb Z}$ and ${\mathbb Q}$ can be chosen to enforce
the normalization condition $\int\! d\rho(x) = \int\! d\rh(y) =1$.

In the $\beta\to\infty$ limit, we adopt the Ansatz
(\ref{BetaInftyAnsatz}) for $\rho(x)$ and $\rh(y)$ and keep
$m\beta=\mu$ fixed. We obtain the following free energy:
\begin{eqnarray}
\phi(\mu) & = & \frac{\la p}{\mu}\log\left\{
1+(e^{-2\mu}-1)[\rho_+\rh_- + \rho_-\rh_+]\right\}-\nonumber\\
&&-\frac{\la}{\ka\mu}\sum_{\nu=0}^{\infty}g_{\nu}\log\left\{
1+\frac{1}{2}(e^{-2\mu}-1)[(\rho_++\rho_-)^{\nu}-
(\rho_+-\rho_-)^{\nu}]\right\}-\\
&&-\frac{p}{\mu}\sum_{l=2}^{\infty}v_l\log\left\{
\sum_{n_+,n_0,n_-}\!\!\!\!\!{}'\,\frac{l!}{n_+!n_0!n_-!}
\rh_+^{n_+}\rh_0^{n_0}\rh_-^{n_-}\, e^{-2\mu\min(n_+,n_-)}
\right\}\, ,\nonumber
\end{eqnarray}
the sum $\sum'$ being restricted to the integers $n_+,n_0,n_-\ge 0$
such that $n_++n_0+n_-=l$.
The saddle point equations are 
\begin{eqnarray}
\rh_+ & = & \frac{1}{2{\mathbb Q}}\sum_{\nu=1}^{\infty}f_{\nu-1} A_{\nu}^{-1}
\, [(\rho_++\rho_-)^{\nu-1}+(\rho_+-\rho_-)^{\nu-1}]\, ,
\label{BetaInftyRSB1}\\
\rh_- & = & \frac{1}{2{\mathbb Q}}\sum_{\nu=1}^{\infty}f_{\nu-1} A_{\nu}^{-1}
\, [(\rho_++\rho_-)^{\nu-1}-(\rho_+-\rho_-)^{\nu-1}]\, ,
\label{BetaInftyRSB2}\\
\rho_+ & = & \frac{1}{{\mathbb Z}\, \la}\sum_{l=2}^{\infty} v_l l\,
B_l^{-1}\!\!\!\!\!\!\sum_{n_+>n_-;\,n_0} 
\frac{(l-1)!}{n_+!n_0!n_-!}\, \rh_+^{n_+}\rh_0^{n_0}\rh_-^{n_-}\
\, e^{-2\mu n_-}\,\delta_{n_++n_0+n_-,l-1}\, ,
\label{BetaInftyRSB3}\\
\rho_- & = & \frac{1}{{\mathbb Z}\, \la}\sum_{l=2}^{\infty} v_l l\, 
B_l^{-1}\!\!\!\!\!\!\sum_{n_->n_+;\,n_0} 
\frac{(l-1)!}{n_+!n_0!n_-!}\, \rh_+^{n_+}\rh_0^{n_0}\rh_-^{n_-}
\, e^{-2\mu n_+}\, \delta_{n_++n_0+n_-,l-1}\, ,
\label{BetaInftyRSB4}
\end{eqnarray}
where
\begin{eqnarray}
A_{\nu} & = & 1+\frac{1}{2}(e^{-2\mu}-1)[(\rho_++\rho_-)^{\nu}-
(\rho_+-\rho_-)^{\nu}]\, ,\\
B_l &= & \sum_{n_+,n_0,n_-}\frac{l!}{n_+!n_0!n_-!}
\rh_+^{n_+}\rh_0^{n_0}\rh_-^{n_-}\,e^{-2\mu\min(n_+,n_-)}\,
\delta_{n_++n_0+n_-,l}\, ,\\
{\mathbb Q} & = & \sum_{\nu = 1}^{\infty} f_{\nu-1}A_{\nu}^{-1}\,
,\\
{\mathbb Z} & = & \frac{1}{\la}
\sum_{l=2}^{\infty} v_l l\, B_l^{-1}\sum_{n_+,n_0,n_-} 
\frac{(l-1)!}{n_+!n_0!n_-!}\, \rh_+^{n_+}\rh_0^{n_0}\rh_-^{n_-}\
\, e^{-2\mu \min(n_+,n_-)}\,\delta_{n_++n_0+n_-,l-1}\, .
\end{eqnarray}

It is interesting to consider some particular asymptotics of the above
results. By taking the limit $\mu\to 0$ we recover the replica
symmetric energy (\ref{RSEnergy}) and the saddle point equations
(\ref{BetaInfty1})-(\ref{BetaInfty4}). In the $\mu \to\infty$ limit
we have $\phi(\mu)\mu\to \phi_{\infty}(\rho,\rh)$. Notice that,
from Eq. (\ref{Complexity}), we get $\phi_{\infty}(\rho,\rh) =
-\Sigma(0)$, $\Sigma(0)$ being the zero-energy complexity.
The explicit expression for this quantity is
\begin{eqnarray}
\phi_{\infty}(\rho,\rh) & = & \la p\,\log [1-(\rho_+\rh_-+\rho_-\rh_+)]-
p\sum_{l=2}^{\infty}v_l\log\{(1-\rh_+)^l+(1-\rh_-)^l-\rh_0^l\}-
\nonumber\\
&&-\frac{\la}{\ka}\sum_{\nu=0}^{\infty}g_{\nu}\log\left\{
1-\frac{1}{2}[(\rho_++\rho_-)^{\nu}-(\rho_+-\rho_-)^{\nu}]\right\}\, , 
\label{FreeBetaInfty_oInfty}
\end{eqnarray}
whose minimization yields the following saddle point equations:
\begin{eqnarray}
\rh_+ & = & \frac{1}{2{\mathbb Q}}\sum_{\nu=1}^{\infty}f_{\nu-1} A_{\nu}^{-1}
\, [(\rho_++\rho_-)^{\nu-1}+(\rho_+-\rho_-)^{\nu-1}]\, ,
\label{BetaInftyRSB1_oInfty}\\
\rh_- & = & \frac{1}{2{\mathbb Q}}\sum_{\nu=1}^{\infty}f_{\nu-1} A_{\nu}^{-1}
\, [(\rho_++\rho_-)^{\nu-1}-(\rho_+-\rho_-)^{\nu-1}]\, ,
\label{BetaInftyRSB2_oInfty}\\
\rho_+ & = & \frac{1}{{\mathbb Z}\, \la}\sum_{l=2}^{\infty} v_l l\,
B_l^{-1}[(1-\rh_-)^{l-1}-(1-\rh_+-\rh_-)^{l-1}]\, ,
\label{BetaInftyRSB3_oInfty}\\
\rho_- & = & \frac{1}{{\mathbb Z}\, \la}\sum_{l=2}^{\infty} v_l l\,
B_l^{-1}[(1-\rh_+)^{l-1}-(1-\rh_+-\rh_-)^{l-1}]\, ,
\label{BetaInftyRSB4_oInfty}
\end{eqnarray}
with
\begin{eqnarray}
A_{\nu} & = & 1-\frac{1}{2}[(\rho_++\rho_-)^{\nu}-
(\rho_+-\rho_-)^{\nu}]\, ,\\
B_l &= & (1-\rh_+)^l+(1-\rh_-)^l-(1-\rh_+-\rh_-)^l\, .
\end{eqnarray}

We look for a solution of Eqs. (\ref{BetaInftyRSB1})-(\ref{BetaInftyRSB4})
which is the analytic continuation of the ``physical'' one identified in
the previous Section for $\mu=0$. Such a solution exists 
in some interval $\mu_1(p)<\mu< \mu_2(p)$. 
For $p<p^*$ no physical solution
exists for any value of $\mu$. 
For $p^*<p<p_d$, $0=\mu_1(p)<\mu_2(p)$ and $\phi(\mu)$
is a monotonously increasing function between 
$\mu_1(p)$ and  $\mu_2(p)$. A physical
solution exists but we cannot associate to it any ``well-behaved''
complexity. 
Above $p_d$ we have $0<\mu_1(p)<\mu_2(p) =\infty$ 
and a ``well-behaved'' complexity can be computed by Legendre-transforming
$\mu\phi(\mu)$\footnote{The situation around $p_d$
is more complicate than the one we described. This is an
artifact of the variational approximation we adopted for computing the
1RSB free energy. Here is a sketch of what happens. At $p\approx 0.419$
a maximum of $\phi(\mu)$, which is still defined between 
$0$ and $\mu_2(p)<\infty$, appears. At $p\approx 0.424$ the
function $\phi(\mu)$ breaks down in two branches: a small $\mu$
(defined between $0$ and $\mu_1(p)>0$), and a large
$\mu$ (defined between $\mu_1(p)$ and $\mu_2(p)<\infty$)
continuation. This second branch has a maximum for some $\mu^*$.
At $p\approx 0.42715$, $\mu_2(p)\to\infty$. 
This threshold can be computed by studying the asymptotic problem
defined by Eq. (\ref{FreeBetaInfty_oInfty}), whose physical solution
is the saddle point lying on the $\rho_++\rho_-=1$ line. 
Finally, at $p=p_d\approx 0.429440$, the small $\mu$ branch
disappears.}, cf. Eq. (\ref{Complexity}).
The complexity $\Sigma(\epsilon)$ is non-zero between $\epsilon_s$ and
$\epsilon_d$. At $p = p_c$ the static energy $\epsilon_s$ vanishes:
more than one codeword (more precisely, about $\exp\{N\Sigma(0)\}$
codewords) is consistent with the received 
message\footnote{Once again, because of the variational approximation
we made in computing $\phi(\mu)$, we obtain $\epsilon_s=0$ above
$p>p'_c\approx 0.48697$.}.
%
%************************************************************************
%
\subsubsection{Beyond the factorized Ansatz}
\label{BeyondSection}

The general one-step replica symmetry breaking order
parameter  \cite{MonassonRSB} is
\begin{eqnarray}
\l(\vs)  =  \int\!\!DQ[\rho]\,\prod_{{\cal G}=1}^{n/m}
\left[\int\!\! d\rho(x)\, \frac{e^{\beta x\sum_{a\in {\cal G}}\sigma^a}}
{(2\cosh\beta x)^m}\right]\, ,\;\;\;\;\;
\lh(\vs)  = \int\!\!D\Qh[\rh]\,\prod_{{\cal G}=1}^{n/m}
\left[\int\!\! d\rh(y)\, \frac{e^{\beta y\sum_{a\in {\cal G}}\sigma^a}}
{(2\cosh\beta y)^m}\right]\, .\nonumber\\
\label{1RSBAnsatz}
\end{eqnarray}
The saddle point equations for functional order parameters
$Q[\rho]$ and $\Qh[\rh]$ are given in the next Section for a general
channel, cf. Eqs. (\ref{GeneralRSB_1}), (\ref{GeneralRSB_2}).

In the previous Section we used a quasi-factorized Ansatz
of the form:
\begin{eqnarray}
Q[\rho] = (1-p)\delta[\rho-\delta_{\infty}]
+p\,\delta[\rho-\rho_0]\, ,\;\;\;\;
\Qh[\rh] = \delta[\rh-\rh_0]\, ,\label{QuasiFactorized}
\end{eqnarray}
where $\delta[\cdot]$ is a functional delta function, and 
$\delta_{\infty}(x)$ is the ordinary Dirac delta centered at $x=+\infty$.
This Ansatz does not satisfy the saddle point equations
(\ref{GeneralRSB_1}), (\ref{GeneralRSB_2}), but yields very good 
approximate results.

Some exact results\footnote{A.M. thanks M.~M\'ezard and R.~Zecchina 
for fruitful suggestions on this topic \cite{MarcRiccardo}.} 
(within an 1RSB scheme) can be obtained by
writing the general decomposition
\begin{eqnarray}
Q[\rho] = u\, Q_s[\rho]+(1-u)\, Q_a[\rho]\, ,\;\;\;\;
\Qh[\rh] = \uh\, \Qh_s[\rh]+(1-\uh)\, \Qh_a[\rh]\, ,
\end{eqnarray}
where $Q_s[\rho]$ and $\Qh_s[\rh]$ are concentrated on the subspace of
symmetric distributions (for which $\rho(x) = \rho(-x)$, $\rh(y) = \rh(-y)$),
while $Q_a[\rho]$ and $\Qh_a[\rh]$ have zero weight on this 
subspace. Using this decomposition in Eqs. 
(\ref{GeneralRSB_1}), (\ref{GeneralRSB_2}), we get, for the BEC,
a couple of equations for $u$ and $\uh$, which are identical to the
replica symmetric ones, cf. Eq. (\ref{SaddleFerro}).

The meaning of this result is clear. For $p>p_d$ the system decompose 
in two parts. There exists a {\it core} which the iterative 
algorithms are unable to decode, and is completely glassy. This part 
is described by the functionals $Q_s[\rho]$ and $\Qh_s[\rh]$. The rest of
the system (the {\it peripheral} region) can be decoded by the
belief propagation algorithm and, physically, is strongly magnetized.
This corresponds to the functionals $Q_a[\rho]$ and $\Qh_a[\rh]$
(a more detailed study shows that the asymmetry of $\rho$ and $\rh$ is,
in this case, typically positive). 
%
%************************************************************************
%
\section{Calculations: the general channel}
\label{GeneralAppendix}

In this Appendix we give some details of the replica calculation for a 
general noisy channel (i.e. for a general distribution $p(h)$ 
of the random fields). In contrast with the BEC case, cf Eqs. 
(\ref{BetaInftyAnsatz}),
the local field distributions do not have a simple form even the 
zero temperature limit. 
Therefore our results are mainly based on a numerical solution of the saddle 
point equations.

\subsection{Finite temperature}
\label{FiniteTemperatureSection}

The one-step replica symmetry breaking Ansatz is given in 
Eqs. (\ref{1RSBAnsatz}).
Inserting in Eq. (\ref{Action}) and taking the $n\to 0$ limit,
we get $S[\l,\lh]=n\phi[Q,\Qh]+O(n^2)$, with
\begin{eqnarray}
\phi[Q,\Qh] & = &\frac{\la}{m}\int\!DQ[\rho]\int\! D\Qh[\rh]\,
\log\left\{\int\! d\rho(x)\int\! d\rh(y) \, [1+\tb(x)\tb(y)]^m\right\}-
\nonumber\\
&&-\frac{\la}{\ka m}\sum_{k=3}^{\infty}c_k
\int\! \prod_{i=1}^k DQ[\rho_i]\log\left\{
\int\! \prod_{i=1}^kd\rho_i(x_i)[1+\tb\tb(x_1)\cdots\tb(x_k)]^m\right\}-
\nonumber\\
&&-\frac{1}{m}\sum_{l=2}^{\infty}v_l\int\!\prod_{i=1}^l D\Qh[\rh_i]\,
\left\<\log\left\{\int\!\prod_{i=1}^l d\rh_i(y_i) 
{\mathbb F}_{l+1}(\frac{\zh h}{\beta},y_1,\dots,y_l)^m\right\}\right\>_h-\nonumber\\
&&-\<\log\cosh(\zh h)\>_h+\frac{\la}{\ka}\log(1+\tb)\, ,\label{Free1RSB}
\end{eqnarray}
where we used the shorthands $\tb(x) = \tanh(\beta x)$, $\tb = \tanh(\beta)$,
and defined 
\begin{eqnarray}
{\mathbb F}_{n}(y_1,\dots,y_n) \equiv \prod_{i=1}^n(1+\tb(y_i))+
\prod_{i=1}^n(1-\tb(y_i))\, .
\end{eqnarray}
The saddle point equations are
\begin{eqnarray}
Q[\rho] & = & \frac{1}{\la}\sum_{l=2}^{\infty}v_l l
\int\!\!dp(h)\int\!\prod_{i=1}^{l-1}D\Qh[\rh_i]\,\,
\delta[\rho-\rho^{(l)}_h[\rh_1,\dots,\rh_{l-1}]]\, ,\label{GeneralRSB_1}\\
\Qh[\rh] & = & \frac{1}{\ka}\sum_{k=3}^{\infty}c_k k
\int\!\prod_{i=1}^{k-1}DQ[\rho_i]\,\,
\delta[\rh-\rh^{(k)}[\rho_1,\dots,\rho_{k-1}]]\, ,\label{GeneralRSB_2}
\end{eqnarray}
where $\delta[\dots]$ denotes the functional delta function, and the
$\rho^{(l)}_h[\dots]$, $\rh^{(k)}[\dots]$ are defined as follows:
\begin{eqnarray}
\frac{\rho^{(l)}_h(x)}{(2\cosh\beta x)^m} & = & \frac{1}{\cal Z}
\int\!\prod_{i=1}^{l-1}\frac{d\rh_i(y_i)}{(2\cosh\beta y_i)^m}\,\,\,
\delta(x-\frac{\zh h}{\beta}-y_1-\dots-y_{l-1})\, ,\label{GeneralRSB_3}\\
\rh^{(k)}(y) & = & \int\!\prod_{i=1}^{k-1}d\rho_i(x_i)\,\,\, 
\delta\left[y-\frac{1}{\beta}\,\atanh[\tb\cdot\tb(x_1)\cdots\tb(x_{k-1})]
\right]\, .\label{GeneralRSB_4}
\end{eqnarray}
These equations can be solved numerically using the population dynamics 
algorithm of Ref. \cite{MezardParisiBethe}. Some outcomes of this 
approach are reported in Sec. \ref{CompleteFiniteBeta}.

\subsubsection{The random linear code limit}
\label{RLC_FiniteTemp_App}
An alternative to this numerical approach consists in considering 
a regular $(k,l)$ code and looking 
at the $k,l\to\infty$ limit with fixed rate $R=1-l/k$.
The leading order results are given in Sec. \ref{RLCbetaSection}.
Here we give the form of the functional order parameters in this limit.

In the ferromagnetic and paramagnetic phases the order parameter 
is replica symmetric: 
\begin{eqnarray}
Q_{P/F}[\rho]=\int\!d\rho_{P/F}(x)\delta[\rho-\delta_x]\, ,\;\;\;\;
\Qh_{P/F}[\rh]=\int\!d\rh_{P/F}(y)\delta[\rh-\delta_y]\, ,
\end{eqnarray}
where $\delta_x$ is a delta function centered in $x$. Moreover we have
$\rho_{P}(x) = p(x)$, $\rh_P(y)=\delta(y)$, and
$\rho_{F}(x) = \rh_F(x) = \delta_{\infty}(x)$. Using these
results in Eq. (\ref{Free1RSB}) we get the paramagnetic and
ferromagnetic free energies, Eqs. (\ref{FreeParaRLC}), (\ref{FreeFerroRLC}).

In the spin glass phase the functionals $Q[\rho]$ and $\Qh[\rh]$ are
non-trivial, although very simple:
\begin{eqnarray}
Q_{SG}[\rho]=\int\! dp(h)\, \delta[\rho-\rho_{SG}^{(h)}]\, ,\;\;\;\;
\Qh_{SG}[\rh]= \delta[\rh-\rh_{SG}]\, , \label{SGRLC0}
\end{eqnarray}
where
\begin{eqnarray}
\rho_{SG}^{(h)}(x) & = & \frac{1}{Z^{(h)}}\sum_{q=0}^{l-1}C^{(h)}_q
\delta(x-\zh h/\beta-2q+l-1)\, ,\label{SGRLC1}\\
\rh_{SG}(y) & = & \frac{1}{2}\delta(y-1)+\frac{1}{2}\delta(y+1)\, ,
\label{SGRLC2}
\end{eqnarray}
and
\begin{eqnarray}
C^{(h)}_q = \frac{1}{2^{l-1}}
\left(\begin{array}{c}l-1\\q\end{array}\right)
[2\cosh(\zh h+\beta(2q-l+1))]^m\, .\label{SGRLC3}
\end{eqnarray}
Substituting in Eq. (\ref{Free1RSB}), we get the spin glass free energy
(\ref{FreeSGRLC}). Notice that the order parameters (\ref{SGRLC1}) and
(\ref{SGRLC2}) can be used to compute the first correction
to the $k,l\to\infty$ limit. For an example of such a calculation we refer
to App. \ref{RLCAppZeroTemp}.
%
%*****************************************************
%
\subsection{Zero temperature}
\label{GeneralZeroTemperature}

In this Appendix we compute the number of metastable states having a 
fixed overlap with a random 
configuration\footnote{Notice that such states are not necessarily 
stable with respect to moves which change their overlap with 
$\us^{\rm out}$.} $\us^{\rm out}$.
The dynamical and statical thresholds for the BSC
can be deduced from the results of this computation, cf. Sec. 
\ref{ZeroTemperatureSection}. The generalization to other statistical 
models for the noisy channel is straightforward (but slightly
cumbersome from the point of view of notation).

In order to study the existence of metastable states,
we consider the constrained partition function:
\begin{eqnarray}
Z(q;\us^{\rm out}) = \sum_{\us} e^{-\beta H_{\rm exch}(\us)}
\delta(Nq-\sum_{i=1}^N\sigma^{\rm out}_{i}\sigma_i)\, ,
\end{eqnarray}
where the received bits $\sigma^{\rm out}_i$ are i.i.d. quenched 
variables: $\sigma^{\rm out}_i=+1$ ($-1$) with probability 
$1-p$ (respectively $p$).
We introduce $m$ ``real'' weakly coupled replicas of the system:
\begin{eqnarray}
Z_m(q;\us^{\rm out}) = \int_{-i\infty}^{+i\infty}\prod_{a=1}^m
\beta \frac{dh_a}{2\pi} \, e^{-Nq\sum_a h_a}
\sum_{\{\us^a\}}e^{-\beta \sum_{a=1}^m H_{\rm exch}(\us^a)+
\beta\sum_{a=1}^m\sum_{i=1}^Nh_a\sigma^{\rm out}_{i}\sigma^a_i}\, .
\end{eqnarray}
For a general channel we should look at the likelihood rather than
at the overlap.

We make the hypothesis of symmetry among the $m$ coupled replicas.
In particular we use the same value of the Lagrange multiplier for  
all of them: $h_a=h_0/\beta m$. We are therefore led to compute
\begin{eqnarray}
\phi(m;h_0) = -\lim_{n\to 0}\frac{1}{n}\log
\overline{\tilde{Z}_m(h_0;\us^{\rm out})^{n/m}}\, ,
\end{eqnarray}
where 
\begin{eqnarray}
\tilde{Z}_m(h_0;\us^{\rm out}) = 
\sum_{\{\us^a\}}e^{-\beta \sum_{a=1}^m H_{\rm exch}(\us^a)+
(h_0/m)\sum_{a=1}^m\sum_{i=1}^N\sigma^{\rm out}_{i}\sigma^a_i}\, .
\end{eqnarray}
Next we take the zero temperature limit  keeping $m\beta=\mu$ fixed.
With a slight abuse of notation, we have $m\phi(m;h_0)\to \mu\phi(\mu;h_0)$.
The entropy of metastable states, cf. Eq. (\ref{ConfEntropy}), is obtained 
as the Legendre transform of $\mu\phi(\mu;h_0)$:
\begin{eqnarray}
\Sigma_p(\epsilon,q) = \mu\epsilon-h_0 q-\mu\phi(\mu;h_0)\, ,
\end{eqnarray}
with $\epsilon = \partial_{\mu}[\mu\phi(\mu;h_0)]$ and 
$q = -\partial_{h_0}[\mu\phi(\mu;h_0)]$.

The replica expression for $\phi(\mu;h_0)$ is easily obtained by taking the 
zero temperature limit on the results of Sec. \ref{FiniteTemperatureSection}.
The free energy reads (for sake of simplicity we write it for a
regular $(k,l)$ code, the generalization is trivial by making use of
Eq. (\ref{Free1RSB})):
\begin{eqnarray}
\mu\phi[Q,\Qh] & = &l\int\!DQ[\rho]\int\! D\Qh[\rh]\,
\log\left\{1+\int\! d\rho(x)\int\! d\rh(y) \,\theta(-xy)
[e^{-2\mu\min(|x|,|y|)}-1]\right\}-
\nonumber\\
&&-\frac{l}{k}
\int\! \prod_{i=1}^k DQ[\rho_i]\log\left\{1+
\int\! \prod_{i=1}^kd\rho_i(x_i)\theta(-x_1\cdots x_k)
[e^{-2\mu\min(1,|x_1|,\dots,|x_k|)}-1]\right\}-
\nonumber\\
&&-\int\!\prod_{i=1}^l D\Qh[\rh_i]\,\sum_{\sigma^{\rm out}}p_{\sigma^{\rm out}}
\log\left\{\int\!\prod_{i=1}^l d\rh_i(y_i) 
\exp[-2\mu{\mathbb E}_{\sigma^{\rm out}}(y_1\dots y_l)]\right\}-\nonumber\\
&&-h_0\, ,\label{ZeroTFree1RSB}
\end{eqnarray}
where $p_{\sigma^{\rm out}}=1-p$ for $\sigma^{\rm out}=+1$, and
$p_{\sigma^{\rm out}}=p$ for $\sigma^{\rm out}=-1$ and
\begin{eqnarray}
{\mathbb E}_{\sigma}(y_1,\dots,y_l) = 
\min\left[\sum_{i:y_i\sigma<0}|y_i|;\, h_0/\mu+\sum_{i:y_i\sigma>0}|y_i|\right]
\, .
\end{eqnarray}
The saddle point equations become in this limit
\begin{eqnarray}
Q[\rho] & = & \frac{1}{\la}\sum_{l=2}^{\infty}v_l l
\sum_{\sigma^{\rm out}}p_{\sigma^{\rm out}}
\int\!\prod_{i=1}^{l-1}D\Qh[\rh_i]\,\,
\delta[\rho-\rho^{(l)}_{\sigma^{\rm out}}[\rh_1,\dots,\rh_{l-1}]]\, ,
\label{ZeroTRSB_1}\\
\Qh[\rh] & = & \frac{1}{\ka}\sum_{k=3}^{\infty}c_k k
\int\!\prod_{i=1}^{k-1}DQ[\rho_i]\,\,
\delta[\rh-\rh^{(k)}[\rho_1,\dots,\rho_{k-1}]]\, .\label{ZeroTRSB_2}
\end{eqnarray}
The functionals $\rho^{(l)}_{\sigma^{\rm out}}[\dots]$, 
$\rh^{(k)}[\dots]$ are defined as follows:
\begin{eqnarray}
\rho^{(l)}_{\sigma^{\rm out}}(x) & = & \frac{1}{\cal Z}
\int\!\prod_{i=1}^{l-1}d\rh_i(y_i)\,\,\,
e^{\mu|x|-\mu\sum_i|y_i|}\,\,\,
\delta(x-(h_0/\mu)\sigma^{\rm out}-y_1-\dots-y_{l-1})\, ,
\label{ZeroTRSB_3}\\
\rh^{(k)}(y) & = & \int\!\prod_{i=1}^{k-1}d\rho_i(x_i)\,\,\, 
\delta\left[y-{\rm sign}(x_1\cdots x_{k-1})\min(1,|x_1|,\dots,|x_{k-1}|)
\right]\, .\label{ZeroTRSB_4}
\end{eqnarray}
\subsubsection{The random linear code limit}
\label{RLCAppZeroTemp}

Here we consider the large $k$, $l$ limit for the zero temperature
free energy $\phi(\mu;h_0)$. While the leading order can be obtained 
through elementary methods, cf. Sec. \ref{ZeroTRLC}, the next-to-leading
order (which is required for obtaining a non-zero dynamic threshold)
must be computed within the replica formalism presented in the previous 
Section.

We will take advantage of the fact that it is sufficient to know
the saddle point order parameters to the leading order, in order to compute
the free energy to the next-to-leading order. 
It is easy to check that, for $k,l\to\infty$ we have
\begin{eqnarray}
Q[\rho]=(1-p)\, \delta[\rho-\rho_+]+ p\, \delta[\rho-\rho_-]\, ,\;\;\;\;
\Qh_{SG}[\rh]= \delta[\rh-\rh_0]\, ,
\end{eqnarray}
where $\rh_0(y) = (1/2)\delta(y-1)+(1/2)\delta(y+1)$ and
\begin{eqnarray}
\rho_{\sigma}(x) & = & \frac{1}{Z_{\sigma}}\sum_{q=0}^{l-1}C^{\sigma}_q
\delta(x-(h_0/\mu)\sigma-2q+l-1)\, ,\\
C^{\sigma}_q & = & \frac{1}{2^{l-1}}
\left(\begin{array}{c}l-1\\q\end{array}\right)
\exp\{|-\mu(l-1)+2\mu q+h_0\sigma|\}\, .
\end{eqnarray}
These expressions can be obtained by taking the zero temperature limit
of Eqs. (\ref{SGRLC0})-(\ref{SGRLC3}).

Substituting these solutions into Eq. (\ref{ZeroTFree1RSB}) we get
\begin{eqnarray}
\mu\phi(\mu;h_0) & = & -(1-R)\log\left(\frac{1+e^{-2\mu}}{2}\right)
-\log(1+e^{-2h_0})-h_0-\label{Largekl}\\
&&-(1-R)\tanh \mu[(1-2p)\tanh h_0]^k+f_l(\mu)(\cosh\mu)^{-l}+
\nonumber\\
&&+O(\tanh h_0^{2k},(\cosh\mu)^{-2l})\, ,\nonumber
\end{eqnarray}
where we defined
\begin{eqnarray}
f_l(\mu)=\left\{\begin{array}{ll}
2^{-l}\sum_{n=(l+1)/2}^l\left(\begin{array}{c} l\\n\end{array}\right)
e^{\mu(l-2n)} &\mbox{for $l$ odd,}\\
\\
2^{-l}\sum_{n=l/2}^l\left(\begin{array}{c} l\\n\end{array}\right)
e^{\mu(l-2n)}-2^{-l-1}\left(\begin{array}{c} l\\l/2\end{array}\right) 
&\mbox{for $l$ even.}\end{array}\right.
\end{eqnarray}
For $l\to\infty$, $f_l(\mu)\approx (2\pi\mu^2)^{-1/2}$.
It is easy to check that Legendre transform of the first three terms of
Eq. (\ref{Largekl}) gives the elementary result (\ref{SigmaRLC}).
Subsequent terms give the leading corrections.

%
%***************************************
%
\subsubsection{A variational calculation} 
\label{VariationalAppendix}

The zero temperature equations simplify in the limit $\mu
\to\infty$, corresponding to vanishing exchange energy. In that case, a finite
value of $q$ is obtained if the magnetic field $h_0$ is kept finite, 
and it can be proved that
the relation $q=\tanh(h_0)$ holds. In this limit, a direct inspection of
the saddle point equations reveals that only the values $\pm (l-1)$ are
possible for the cavity fields $x$, and the values $\pm 1$ for the
$y$'s.
More explicitly, the order parameters $Q[\rho]$ and $\Qh[\rh]$
are supported on distributions of the form
\begin{eqnarray}
\rho(x) = \rho_+\delta(x-l+1)+\rho_-\delta(x+l-1)\, ,\;\;\;\;\;\;
\rh(y) = \rh_+\delta(y-1)+\rh_-\delta(y-1)\,\, .
\label{AnsatzLast}
\end{eqnarray}
The functional order parameter $\Qh[\rh]$, reduces to the 
probability distributions of a single number ${\hat \rho}_+$ representing the
probability of $y=+1$. 

A simple approximation is obtained by using (\ref{AnsatzLast}) and neglecting
the fluctuations of ${\hat \rho}_+$, in the spirit of the
``factorized Ansatz'' of \cite{FranzEtAlExact}. 
This is exact~\footnote{This assertion is true only for even values of $l$,
but actually it is a very good approximation for any value of $l$.} 
for $h_0=0$, where our model reduces to the one analyzed
in \cite{FranzEtAlExact}. 
It can be proved that, for $\mu=\infty$ and $h_0\ne 0$, this approximation
gives the same result as the $k,l\to\infty$ limit, cf. Sec. \ref{ZeroTRLC}.  
For instance in the case of $(k,l)=(6,5)$ we get
$p_c^{var}=0.264$ which coincides with the exact result. 

\begin{figure}
\centerline{\epsfig{figure=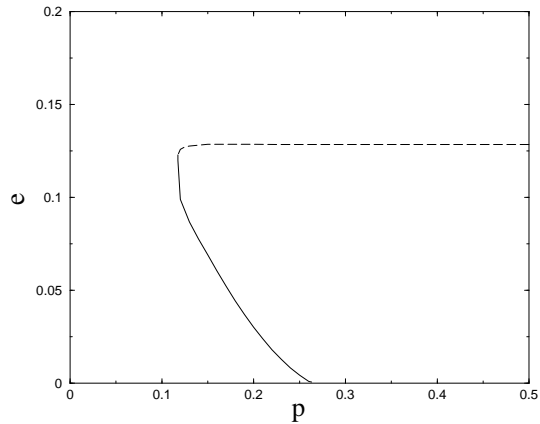,angle=-90,width=0.4\linewidth}}
\caption{The region of metastability as
predicted by the approximated Ansatz (\ref{AnsatzLast}) for the (6,5) code.}
\label{figure1}
\end{figure}

The same form for the functional order parameter can  also be 
used as a variational approximation for $\mu$ finite, although in this
case it is not justified to assume $y=\pm 1$. In Fig. \ref{figure1},
we indicate the region of the $(p,\epsilon)$ plane 
such that $\Sigma_p(\epsilon,1-2p)>0$, as obtained from this
simple approach.
%
%
%************************************************************************
%

\end{document}